# Modelling the Dynamics of Potentially Dangerous Large-Earth Impactors

J. G. Muthini [a*], and G. O. Okeng'o [a]

[a] The Astronomy and Astrophysics Thematic Research Group, Department of Physics, University of Nairobi, Kenya

ABSTRACT

Collisional threats posed by Near-Earth Objects (NEOs) are increasingly being confirmed by National Aeronautics and Space Administration (NASA) and European Space Agency (ESA) sky surveys. Efforts to develop tools to perform modelling, analysis and prediction of possible future impact events are ongoing. The aim of this research is to design a model for Large Earth impact events, describing atmospheric entry, ground impact and post-impact dynamics for impactors large enough to cause harm to the Earth. We present results of Large Earth Impact (LEI) events involving a number of objects with densities ranging between 1000 and 8000 $kg\ m^{-3}$. We use Python code to solve differential equations and perfom calculations on analytical models built into the collision simulation algorithms while feeding relevant physical input in terms of object velocity and diameter. We find that denser impactors have slower rates of decrease in momentum while hitting the atmosphere. The denser impactors which exhibit greater resistance to atmospheric disruption result in more energetic ground impacts events in terms of crater formation, thermal radiation generation, seismic effects, ejecta displacement and airblast effects. We conclude that impacting objects with high masses maintain greater kinetic energies compared to onbjects with less mass thus posing greater collisional threats to the Earth.

## 1. Introduction

Earth has been subjected to impacts by large objects with varying degrees of severity according to paleontological and geological accounts from both near and the distant past. One such account is the Tunguska airburst event on 30th June 1908, which flattened about 2000 km2 of the Siberian Forest according to Zotkin & Tsikulin (1966) (cited in Chyba, Thomas, & Zahnle, 1993). Dating performed on several large impact craters reveal those geological and evolutionary events such as the dinosaur extinction are driven by highly energetic and dynamic impact events. Evidently, there are threats posed by sizable impactors from space to Earth, including existential threats to all lifeforms (Remo & Haubold, 2014).

Sky surveys performed by the National Aeronautics and Space Administration (NASA) and the European Space Agency (ESA) using ground based telescopes assist in tracking, monitoring and cataloging various objects in space. This includes asteroids and comets nudged by the combined gravitational effect of nearby planets and the sun eventually achieving orbital proximity to or even crossing the Earth's orbit, also known as Near-Earth objects (NEOs). Scientists use precise orbit calculations to predict which objects are likely to cross the Earth's orbital path. As of December 2019 approximately 215851 Near-Earth Asteroids have been discovered with 9021 objects having diameters greater than 1 km (NASA, 2019)[1]. These objects are therefore the main focus of most impact monitoring systems.

This realization of the dangers posed by Near-Earth objects has prompted calls from scientific and humanitarian organizations to develop strategic action plans to prepare for, address and mitigate possible future impact threats (Haubold & Nadis, 2014). Scientific organizations like NASA and ESA have made efforts in identification and characterization of NEOs, understanding Earth impact processes and developing proper impact mitigation strategies. These techniques rely hugely on development of quantitative modelling and analysis techniques that provide information on Earth Impact Probability prediction in terms of timing and location, effects and consequences (United States National Science and Technology Council, 2018).

Earth Impact event processes are complex and their analysis requires equally complex computer codes that simulate impact shock waves and their effects in terms of stress and deformation Pierazzo et al. (2008) (Pierazzo, et al., 2008). Computational simulation capabilities however allow for numerical modelling of these impact events using shock or hydrodynamic codes. Currently, the existence of computational tools such as the Asteroid Risk Mitigation and Optimization Research (AMOR) tool have made it possible to calculate impact risk in terms of risk corridors (Rumpf, 2014). With the help of OrbFit software that computes ephemeris of Virtual Impactors (VIs), scientists are capable of predicting distributions of impacts. The ARMOR tool finds the probability distribution of impacts. The ARMOR tool finds the probability distribution of asteroid impact locations corresponding to asteroid ephemeris by utilizing three Python-based modules. These modules are the Solar System Propagator, the Virtual Impactor Finder and the Risk Corridor Calculation.

The Solar System Propagator module computes the total gravitational force in terms of the N-body problem within the solar system. The Sun, planets (including Pluto) and their moons are major bodies involved in calculating asteroid trajectories based on original ephemerides and epochs. The orbital data of these bodies are referenced from a table and the computed trajectories are recorded for later processing. The Virtual Impactor Finder module computes uncertainties in an asteroid's trajectory and their nominal orbit solutions. Finally both the computed orbital locations and their marginal regions of uncertainty are utilized in assessment of future collision risk with the Earth's surface by the Risk Corridor Calculation Module, which are yielded in terms of probabilities.

Although the ARMOR tool has demonstrated accurate probability predictions of a global nature of asteroid threats (after passing validation tests), impact effects on the human population and the environment, which is a crucial tool for quantitative analysis of impact dynamics, is mostly disregarded thereby needing further attention. Therefore the purpose of this research is to investigate the dynamical processes of Large-Earth Impact events by performing computer simulations of potential future impact events.

This paper is arranged as follows: In the theoretical input section we present a review of the equations describing dynamics of atmospheric entry, ground impact and post ground impact events together with the assumptions involved. In the results and discussion section we present the findings of the computer simulation performed a number of objects with different densities. Two other physical inputs, object velocity and diameters, are obtained from two Near Earth asteroids, object 433 *Eros* and (29075) 1950*DA*.

## 2. Theoretical Input

Modelling impact dynamics involves detailed description of mass, momentum and energy conservation equations that account for the behavior of impacting bodies during and after the event. The first section contains a review of atmospheric entry events driven by fluid interaction dynamics. The second section reviews the ground impact effects that engages mechanics of collisions between solid surfaces, shock effects and evolution of impact energy.

### 2.1. Atmospheric Entry Dynamics

Atmospheric entry events are dependent on a number of parameters describing impactor characteristics and the nature of their interaction with Earth's atmosphere in detail. These entry processes are complex and the accuracy of any atmospheric entry model will depend on factors keyed in the

---

model design. However, the key input parameters identified for an entry model include the velocity of the impactor $v_i$, its density $\rho_i$, its diameter $L_i$ and the angle of entry $\theta$ between the impactor's path and the tangential line parallel to the Earth's surface (Collins, Melosh, & Marcus, 2005).

The entry model is built on several important assumptions that simplify the modelling process. The Earth's atmosphere is composed mainly of gases that give it its fluid characteristics. Therefore its constituent gas particles are bound to collide with an approaching impactor, resulting in the deceleration of the object as it traverses towards the bottom reaches of the atmosphere. The opposing drag force that increases exponentially with depth of the atmosphere influences the characteristic deceleration rate of the impactor. According to Archimedes' Principle, this entry has minimum influence on an impactor's shape, energy and momentum as the buoyant force acting on it is proportional to the displaced atmospheric mass.

Assuming our impactor doesn't achieve free fall velocities, the angle subtended by the trajectory that deviates towards a vertical path is ignored for fairly large and massive impactors (greater than 1 $km$ in diameter). The impactor mass is assumed to be constant given that thermal radiation generated by shock does not significantly ablate the mass of the large impactor. Entry acceleration due to the Earth's gravitational pull is also not factored in assuming the impactor velocity is its velocity relative to a massless Earth.

Considering the exponential nature of the Earth's atmospheric density where density increases with depth from the upper atmosphere, we can express this as

$$\rho(z) = \rho_s \exp\left(\frac{-z}{h}\right) \quad (1)$$

where $\rho_s$ is the Earth's atmospheric density just above the Earth's ground surface, taken to be 1 $kg\ m^{-3}$, $z$ the vertical height above the terrestrial ground surface and $h$ the atmospheric scale height given an approximate value of 8000 $m$. The altitude above the Earth's surface at which the atmospheric drag effect is first experienced, known as the Kármán Line, is approximately $1.0 \times 10^5\ m$ also the vertical distance traversed by an impactor before it hits the ground.

Assuming the disruption of the impactor ensues just as it enters the atmosphere, the resulting atmospheric shock pressure induces axial stresses parallel to but opposite the direction of motion of the leading face. Given an impactor roughly shaped as a right cylinder, the pressure acting on its front and rear faces will be different, with a near vacant pressure acting on the rear part. This rear pressure is assumed to be less than the pressure pelting on the sides. As a result, the impactor experiences a slight but incessant increase in its transverse diameter (Chyba, Thomas, & Zahnle, 1993). This creates a Pancake effect, where the exponential increase in opposing atmospheric pressure causes an increase in the impactor diameter as it plunges deeper into the atmosphere (Collins, Melosh, & Marcus, 2005). This Pancake model can be described by the differential equation

$$\frac{d^2L(z)}{dz^2} = \frac{C_d \rho(z)}{L(z)\rho_i(z)\sin^2\theta} \quad (2)$$

where $L(z)$ is the impactor diameter as a function of altitude, $C_d$ is the atmospheric drag coefficient assigned a maximum value of 2 for a (flattening) right cylinder while $\rho_i$ is the density of the impactor at a given altitude expressed as

$$\rho_i(z) = \frac{3M_i}{4\pi L(z)} \quad (3)$$

with $M_i$ as the mass of the impactor which is constant in the entire entry process.

The energy Ei of an impactor with a near-spherical shape (also adopted in this model) is given by

$$E_i(z) = \frac{\pi}{12}\rho_i(z)L_i^3(z)v_i^2 \quad (4)$$

According to Collins, Melosh, & Marcus, (2005), the impactor velocity may be obtained from the solution to the differential equation

$$\frac{dv_i(z)}{dz} = \frac{3\rho C_d v_i(z)}{4\rho_i L(z)\sin\theta} \quad (5)$$

while an estimate of the elastic strength $Y_i$ of the impactor can be given by

$$\log Y_i = 0.0624\sqrt{\rho_i(z)} + 2.107 \quad (6)$$

The stagnation pressure $P_s$ acting on the leading face of the impactor following Bernoulli's principle can be obtained from (Collins et al., 2005)

$$P_s(z) = \rho(z)(v_i(z))^2 \quad (7)$$

Since the intensity of the pressure wave acting axially on an impactor increases with distance traversed by the impactor, according to Equations 1, 3 and 6 the yield strength decreases with decreasing density. At a given height $z_{br}$, known as the altitude of breakup, the stagnation pressure must equal the elastic strength of the impactor where $P_s(z_{br}) = Y_i(z_{br})$. Below the breakup height, the rate of diameter expansion increases with further descent towards the ground surface (Chyba, Thomas, & Zahnle, 1993). The ratio of the impactor diameter at a given height $L(z)$ to the initial diameter before entry ($L_i$) is therefore expressed as

$$F_p = L(z)/L_o \quad (8)$$

Another assumption is made that the large impactor is not subjected to airburst events within the atmosphere as the impactor remnants remain relatively intact after breakup (Chyba, Thomas, & Zahnle, 1993). Therefore the atmospheric entry model mimics the evolution of impactor properties (energy and momentum) until ground impact ensues.

2.2. Ground Impact Dynamics

Crater Formation and Collapse

The collision dynamics of a large impactor with terrestrial material largely depends on yielded parameters from the atmospheric entry event. This includes impactor momentum and energy and structural properties of colliding bodies. However, modelling of resultant impact crater dimensions relies solely on extrapolation of data from available small-scale cratering experiments, large-scale simulations, nuclear detonation experiments and observation of cratering processes on nearby celestial bodies according to Holsapple & Schmidt (1982) Holsapple & Schmidt (1982), Gault (1974) and Schmidt & Housen (1987) (cited in Collins, Melosh, & Marcus (2005)). The first model obtained from Collins, Melosh, & Marcus (2005) applies scaling rules for estimation of the transient crater diameter, $D_{TC}$, which is the transitional diameter that later collapses and modifies to form the final crater diameter. The scaling law combines target density and atmospheric entry output parameters as shown in the relation below

$$D_{TC} = kG_e^{-0.22}\left(\frac{\rho_f}{\rho_t}\right)L_f^{0.78}\sin^{1/3}\theta \quad (9)$$

where $v_f$ is the final impactor velocity, $\rho_f$ the final impactor density, $\rho_t$ the target density, $\theta$ the angle of inclination of impact trajectory and $G_e$ the surface gravity of the Earth in $m\ s^{-1}$. The contant $k$ depends on the target material where it equals 1.161 for solid rock targets according to Schmidt & Housen (1987) (cited in Collins et al., 2005). The value $\rho_t = 2500\ kg\ m^{-3}$ applies for crystalline targets and $\rho_t = 2750\ kg\ m^{-3}$ for sedimentary targets.

Based on values obtained for $D_{TC}$, Earth based craters are classified into two categories namely (i) Simple craters where $D_{TC} < 3.2\ km$ and (ii) Complex craters where $D_{TC} > 3.2\ km$.

2.2.1. Simple Craters

According to Dence (1965) (as cited in Collins, Melosh, & Marcus (2005)), these craters are characterized by an instinctive morphology where impacts trigger the displacement and excavation of breccia and molten rocks from the transient cavity. These materials are eventually swept back inside along the steep transient crater walls, filling the base of the crater and forming a breccia-melt mixture Grieve, Dence, & Robertson (1977). (Collins, Melosh, & Marcus, 2005) matched observed data to predictions of the model derived by Grieve & Garvin (1984) (cited in Collins, Melosh, & Marcus (2005)) used to estimating simple crater dimensions and obtained a first order approximation of rim to rim final crater diameter $D_{FR}$ yielded by

$$D_{FR} \approx 1.25 D_{TC} \quad (10)$$

while the modelled relation between unbulked breccia lens volume $V_{bl}$ and final crater diameter as observed by Grieve & Garvin (1984) (cited in Collins, Melosh, & Marcus (2005)) is obtained from

$$V_{bl} \approx 0.032 D_{FR}^3 \quad (11)$$

The depth of the transient crater $d_{TC}$ according to Dence (1965) (as cited in Collins, Melosh, & Marcus (2005)) can be estimated using

$$d_{TC} = D_{TC}/2\sqrt{2} \quad (12)$$

while the final crater rim height $h_{FC}$ is obtained from (Collins, Melosh, & Marcus, 2005)

$$h_{FC} = 0.07 \frac{D_{TC}^4}{D_{FR}^3} \quad (13)$$

Therefore, the breccia thickness tbl is estimated as shown below by assuming a characteristic lens-like, parabolic surface at the top and a 10% bulk volume increase due to brecciation process (Collins, Melosh, & Marcus, 2005):

$$t_{bl} = 2.8 \frac{d_{TC} + h_{FC}}{d_{TC} D_{FR}^2} \quad (14)$$

2.2.2. Complex Craters

This class of craters are characterized by their unintuitive morphology according to Dence (1965) (as cited in Collins, Melosh, & Marcus (2005)) where transient crater collapse entails closure of the transient crater by gravitational forces and resistance of post-impact target rocks. According to Croft (1985), McKinnon & Schenk (1985) and Holsapple (1993), restoration of lunar based complex transient craters enable establishment of scaling laws for estimating final rim to rim diameters. Collins, Melosh, & Marcus (2005) apply the equation McKinnon & Schenk (1985)

$$D_{FR} = \frac{1..17 D_{TC}^{1.13}}{D_C^{0.13}} \quad (15)$$

which lies within the approximations of Croft (1985) and Holsapple (1993), where $D_C$ represents the diameter at which craters transition from simple to complex. $D_C$ is usually assigned a value of $3200\ m$ for terrestrial craters. The value substituted for $D_C$ by Collins, Melosh, & Marcus (2005) is $2560\ m$, also implemented in this model.

Estimation of the average complex crater depth $d_{FR}$ applies the relationship between depth and diameter as defined by Herrick, Sharpton, Malin, & Freely (1997). For Venusian craters (cited in Collins, Melosh, & Marcus (2005)) which is reliably consistent with Earth based craters (Pike, 1980). This relation is shown below as

$$d_{FR} = 0.294 D_{FR}^{0.301} \quad (16)$$

For the melt volume estimate, Collins, Melosh, & Marcus (2005) prescribe several conditions that facilitate melt production for their model "based on the results of numerical modeling of the early phase of the impact event (O'Keefe & Ahrens, 1982b; Pierazzo, Vickery, & Melosh, 1997; Pierazzo & Melosh, Melt production in oblique impacts, 2000); and geological observation at terrestrial craters (Grieve & Cintala, 1992). Such impacts involve velocities greater than $\approx 12\ km\ s^{-1}$ and impactor densities comparable to that of the target. The assumption that the melt volume and the transient crater volume are proportional enables computation of a volume of melt estimate using the following equation implemented the model (Collins, Melosh, & Marcus, 2005):

$$V_M = 8.9 \times 10^{12} E \sin\theta \quad (17)$$

where $E$ is the energy of the impact and $\theta$ the angle between impact trajectory and the ground surface. This equation well represents all geological surfaces except ice which is expected to be 10 times more voluminous (Pierazzo, Vickery, & Melosh (1997) cited in Collins, Melosh, & Marcus (2005)). Geological materials close to the Earth's center where temperatures are closer to their melting points are also not considered, especially in cases involving high energy impacts. For low-angle impacts $\theta < 15°$, Equation 17 overestimates the volume by a factor of 2 (Pierazzo & Melosh, 2000). For highly energetic impact events, yielded melt is well mixed with breccia in large complex craters and maintains uniform thickness according to Grieve, Dence, & Robertson (1977) (cited in Collins, Melosh, & Marcus (2005). Collins, Melosh, & Marcus (2005) suggest that an assumption can be made that the crater diameter is similar to the floor diameter. Therefore the ratio of crater melt volume to the base area yields the thickness of melt given by

$$t_M = 4V_M/\pi D_{TC}^2 \quad (18)$$

2.3. Post-Impact Dynamics (Ground Impact)

2.3.1. Thermal Radiation

Impact-induced interaction between the impactor and the ground target leads to compression and melting of constitutive materials triggered by generation of high pressures and temperature at regions closest to the impact site. According to Collins, Melosh, & Marcus (2005) impact pressures high enough melt the impactor and small portions of targeted material involve velocities higher than $12\ km\ s^{-1}$. Impact velocities exceeding $15\ km\ s^{-1}$ lead to vaporization of material characterized by extreme pressures greater than $100\ GPa$ and temperatures above $10000\ K$. Under these conditions, the generated fireball consisting of trapped vapor-radiation mixture expands at a fast rate. High temperatures that ionize the air within the fireball result to enhanced radiation absorption properties. Therefore the confined radiation cannot escape and heat the surrounding regions due to the fireball's temperature-induced opacity. Collins, Melosh, & Marcus (2005) suggest an in depth explanation to this complex process by Glasstone & Dolan (1977). The fireball however is subjected to continual expansion and thereby experiences drops in temperature as the outer confines cover more radii. Upon continual cooling, the surface temparature drops to transparency temperature, $T_*$, according to Zel'dovich & Raizer (1966) (cited in Collins, Melosh, & Marcus, 2005). At this temperature the fireball achieves transparency to radiation which escapes and heats the surrounding regions, initially at high intensity, lasting for some seconds to several minutes. The heating intensity decays as the fireball expands further until emission stops. Nemtchinov, et al. (1998) (cited in Collins, Melosh, & Marcus, 2005)) present the range within which transparency temperature of Earth falls as the visible infrared range $2000 - 3000\ K$. The transparency temperature of Silicate vapor is an approximate value of $6000\ K$ according to Melosh, et al. (1993). Therefore "the limiting factor for terrestrial impacts is the transparency temperature of air surrounding the silicate vapor fireball" (Collins, Melosh, & Marcus, 2005).

Melosh, et al. (1993) and Nemtchinov, et al. (1998) carried out numerical simulations that suggest the fireball radius at the time of transparency is $10 - 15$ times the diameter of the impactor. Collins, Melosh, & Marcus (2005) apply the yield scaling technique to estimate the radius of the fireball $R_{fb}$ from the impact energy $E_f$ given that the impact velocity exceeds $15\ km\ s^{-1}$ using the equation

$$R_{fb} = 0.002 E_f^{1/3} \quad (19)$$

where the constant 0.002 was computed by dividing $R_{fb} = 13L$ by $E_f$ involving an impactor with ground impact velocity $20\ km\ s^{-1}$ and density $2700\ kg\ m^{-3}$. The time of maximum thermal radiation Tmax is estimated under the assumption that the velocity of impact is equal to the initial rate fireball expansion (Collins, Melosh, & Marcus, 2005)

$$T_{max} = \frac{R_F}{v_f} \quad (20)$$

where $v_f$ is the impactor velocity upon ground impact.

Thermal radiation generated is dependent on the fraction of impact energy that generates thermal energy referred to as the luminous efficiency b. Nemtchinov, et al. (1998) suggest that b relates with the impact velocity by a scaled power law. According to a study by Ortiz, et al. (2000) (cited in Collins,, Melosh, & Marcus, 2005) based on few observational, numerical and experimental results that exist show that b ranges from $10^{-4} - 10^{-2}$. Collins, Melosh, & Marcus (2005) obtained a first-order estimate of $3 \times 10^{-3}$ ignoring the undefined constraints for impacts in the hypervelocity range, also adopted in the present model. Therefore thermal exposure per unit area at a distance r from the fireball radius is dependent on the luminous efficiency assuming a hemispherical fireball as demonstrated below

$$\alpha = \frac{0.003 E_F}{2\pi r^2} \quad (21)$$

Fireball expansion from the moment radiation at maximum intensity is released at transparency temperature to the time all trapped radiation has escaped and ceased occurs within a limited time period. This period known as duration of irradiation is obtained by dividing the from Equation 21 at the fireball radius $r = R_{fb}$ by the radiant energy flux (Collins, Melosh, & Marcus, 2005):

$$T_{ir} = \frac{\beta E_F}{2\pi R_{fb}^2 \sigma T_*^4} \quad (22)$$

where the Stefan-Boltzmann constant $\sigma = 5.67 \times 10^{-8} W\ m^{-2}\ K^{-4}$ and the transparency temperature $T_* = 3000K$.

Thermal radiation at distances where part of the fireball is obstructed below the Earth's horizon can be modified by first obtaining the fraction of the fireball below the horizon $f$

$$f = (1 - \cos \Delta) R_E \quad (23)$$

where $\Delta$ is given by $r = R_E$ and $R_E$ is the Earth's radius. $f$ is then used to obtain the area of the fireball above the horizon $j$ that bathes the region at distance r using

$$j = \frac{2}{\pi}\left(\delta - \frac{f}{R_{fb}} \sin \delta\right) \quad (24)$$

where $\delta$ is half the angle of the firebal region visible above the horizon given as

$$\delta = \cos^{-1}\left(\frac{f}{R_{fb}}\right) \quad (25)$$

Therefore the corrected thermal exposure is given by $j\alpha$. However, this case does not consider the extinction and refraction of radiation rays directly above the horizon (Collins, Melosh, & Marcus, 2005). Also cases involving different atmospheric conditions that affect absorption properties of the surrounding atmosphere are not considered.

Ignition of materials exposed to thermal radiation by a given impact energy depends on thermal exposure at the region of interest and the duration of irradiation (Collins, Melosh, & Marcus, 2005). Through application of various scaling laws, the ignition of materials and sustained burns on the skin for a 1 $Mt$ explosion can be used to compute the thermal exposure required to ignite a material for any given impact energy in Megatons. This relation is given by

$$\Omega_{ig}(E) = \Omega_{ig}(1\ Mt) E_{Mt}^{1/6} \quad (26)$$

where a particular material is ignited (or skin burns sustained) if $j\alpha > \omega_{ig}$ at a given distance $r$ from the site of impact. Thermal exposure values at which materials are ignited or burns sustained during a 1 $Mt$ impact event are as given in Table 1 from Glasstone & Dolan (1977) data (cited in Collins, Melosh, & Marcus, 2005)

| Material | Thermal exposure required to ignite material in a 1 Mt explosion ($\Omega_{ig}(1\ Mt)\ MJ\ m^{-2}$) |
|---|---|
| Clothing | 1.0 |
| Plywood | 0.67 |
| Grass | 0.38 |
| Newspaper | 0.33 |
| Deciduous Trees | 0.25 |
| 3rd degree burns | 0.42 |
| 2nd degree burns | 0.25 |
| 1st degree burns | 0.13 |

Table 1: Thermal exposure required to ignite different materials a 1 $MT$ explosion event from Glasstone and Dolan (1977) (cited in Collins et al. 2005)

### 2.3.2. Seismic Effects

Ground impact generates shock waves at the target site that are propagated along the surface and through the target material. The strength of the waves decay as the radii of ground coverage from the impact site increases leaving only waves of an elastic nature. The waves have similar travel mechanisms as those of Earthquakes although they may differ in structure and source. The energy of the waves are a fraction of the total impact energy, represented by a value known as the seismic efficiency. This fraction is also the seismic energy which is roughly approximated by $E_f \times 10^{-4}$, or one part in ten thousand of the impact energy, (Schultz & Gault, 1975). The magnitude of waves generated $M_s$ is yielded from the seismic energy using the Gutenberg-Richter relation used by Melosh (1989) (cited in Collins, Melosh, & Marcus, 2005) is shown below

$$M_s = 0.67 \log_{10} \mu E_f - 5.87 \quad (27)$$

Where the ground impact energy $E_f$ is in Joules.

Propagation of a seismic wave with its intensity depends on the nature of deformation and wave flow within a given geological region. Based on curve fits on empirical data from ground shaking events performed by Ritcher (1958) (cited in Collins, Melosh, & Marcus, 2005), the relation between the effective seismic magnitude ($M_e$) and the distance from the impact site ($r$) in kilometers is represented using the following equations.
For $r_{km} < 60\ km$:

$$M_e = M_s - 0.0238 r_{km} \quad (28)$$

for $60\ km \leq r_{km} \leq 700\ km$

$$M_e = M_s - 1.1644 - 0.0048 r_{km} \quad (29)$$

and for $r_{km} \geq 700\ km$

$$M_e = M_s - 6.399 - 1.66 \log_{10} \Delta \quad (30)$$

The extent of damage incurred can be estimated using the Modified Mercalli Intensity scale which links the effective seismic magnitude with the damage intensity according to Richter's data on Earthquakes (Ritcher, 1958). This relation is shown in Table 2 below (Collins, Melosh, & Marcus, 2005)

| Ritcher Magnitude | Modified Mercalli Intensity |
|---|---|
| 0 - 1 | - |
| 1 – 2 | I |
| 2 – 3 | I – II |
| 3 – 4 | III – IV |
| 4 – 5 | IV – V |
| 5 – 6 | VI – VII |
| 6 – 7 | VII – VIII |
| 7 - 8 | IX – X |
| 8 - 9 | X – XI |
| 9+ | XII |

Table 2: Estimated Mercalli Intensities for different effective magnitudes

The arrival time of the seismic wave with the largest displacement can be estimated by assuming their velocity within uppersurface crustal rocks is approximately 5 $km\ s^{-1}$. Hence the following equation yields the arrival time of the most energetic wave

$$T_s = \frac{r_{km}}{5} \quad (31)$$

### 2.3.3. Ejecta Deposition

Ground impacts by impactors results in displacement of materials at regions near the pre-impact target surface. Some materials are launched as projectiles that eventually land as deposited ejecta material while the rest are simply uplifted and upturned forming additional portions of the transient crater rim. Regions closest to the rim carry large amounts of upturned materials. This amount then decreases at a fast rate as one moves away from the center of the transient crater. The material within two radii from the transient crater center above the original ground surface is almost entirely displaced material. According to Collins, Melosh, & Marcus (2005) this makes it possible to estimate the ejecta thickness within a given location of interest. Therefore the height of the transient crater rim hTR which is also the point of maximum ejecta thickness decreases at a rate determined by the inverse of the distance cubed as shown below:

$$\tau_{ej} = \frac{h_T}{8}\left(\frac{d_{TC}}{r}\right)^3 \quad (32)$$

According to McGetchin, Settle, & Head, (1973) the power $-3$ gives the best fit for data observations on ejecta thickness decay with increase in distance although different sets of experiments yield values within the range $-2$ to $-3.7$. Collins, Melosh, & Marcus (2005) suggest "a simple volume conservation argument" for estimating the thickness of ejecta at the rim of the transient crater. The total volume of ejected material (including the upturned portion that is part of the transient crater rim) $V_E$ is equated with the volume of the transient crater cavity. Assuming the crater is a paraboloid cavity with diameter to depth ratio $2\sqrt{2}:1$, the volume of excavated material is first obtained from Equation 33

$$V_E = \frac{H_{TR}D_{TR}^3}{8}\int_{D_{TR}/2}^{\infty}\frac{2\pi r dr}{r^3} + \int_{D_{TC}/2}^{D_{TR}/2}2\pi r\left(\frac{4d_{TC}}{D_{TC}^2}r^2 - d_{TC}\right)dr \quad (33)$$

$$= \frac{\pi}{2}\left(h_{TR}D_{TR}^2 + d_{TC}\left[\frac{D_{TR}^4 - D_{TC}^4}{4D_{TC}} - \frac{D_{TR}^2 - D_{TC}^2}{2}\right]\right)$$

where $D_{TR}$ is the rim-to-rim transient crater diameter given by

$$D_{TR} = D_{TC}\sqrt{\frac{h_{TR} + d_{TC}}{d_{TC}}} \quad (34)$$

The transient crater volume is then obtained from

$$V_{TC} = \frac{\pi D_{TC}^3}{16\sqrt{2}} \quad (35)$$

According to Collins et al. (2005) the two equal volumes, $V_E = V_{TC}$, can be rearranged to give the transient crater rim height as $h_{TR} = D_{TC}/14.1$. Substituting this relation in Equation 32 yields the relation

$$\tau_{ej} = \frac{D_{TC}^4}{112r^3} \quad (36)$$

The above estimation of the ejecta thickness assumes absence of sizable ejecta as well as non-displacement of substratal material. Therefore only the lowest possible thickness is obtained. The transient crater diameter is relied upon in Equation 36 to avoid the need for separate thickness equations for both simple and complex final craters. Complex crater rim heights are ideally smaller than those of simple craters owing to their collapse in the later stages of the cratering process. Therefore the thickest portion of ejecta material equal to the rim height of the final crater is given by Equation 13 (from Crater Formation and Collapse) obtained by substituting $r = D_{FR}/2$ in Equation 36.

According to Ahrens & O'Keefe (1978), the launched ejecta material eventually follow trajectories whose times of landing can be determined for a spherical planet. Collins, Melosh, & Marcus (2005) suggest that an estimate of mean arrival time of ejecta bulk can be obtained by assuming all ejecta are launched at angles of $45°$ to the horizontal. Under this assumption the ellipticity of a given ejecta trajectory e is given by

$$e^2 = \frac{1}{2}\left[\left(\frac{v_{ej}^2}{g_E R_E} - 1\right)^2 + 1\right] \quad (37)$$

where $v_{ej}$ is the the ejection velocity obtained from

$$v_{ej}^2 = \frac{2G_E R_E \tan\Delta/2}{1 + \tan\Delta/2} \quad (38)$$

The semi-major axis $a$ of the trajectory given by

$$a = \frac{v_{ej}^2}{2G_E(1-e^2)} \quad (39)$$

According to Ahrens & O'Keefe (1978), given conditions where $v_{ej}/(G_E R_R) \leq 1$ and $-1 < e < 0$. The arrival time of the ejecta $T_{ej}$ within $r \approx 10^4\ km$ (Collins, Melosh, & Marcus, 2005) is given by

$$T_{ej} = \frac{2a^{3/2}}{\sqrt{G_E R_E^2}}\left[2\tan^{-1}\sqrt{\frac{1-e}{1-e}}\tan\frac{\Delta}{4} - \left(\frac{e\sqrt{1-e^2}\sin(\Delta/2)}{1 + e\cos(\Delta/2)}\right)\right] \quad (40)$$

where $G_E$ represents the Earth's surface gravitational acceleration, $R_E$ is the radius of the Earth and the epicentral angle $\Delta = r/R_E$. However if the aforementioned conditions are not satisfied, an identical equation provided by Ahrens & O'Keefe (1978) calculates arrival times greater than an hour. Collins, Melosh, & Marcus (2005). At greater distances, the Earth's rotation and atmospheric conditions influences the distribution of mostly fine ejecta. These complex processes are not implemented by the present model.

Schaller & Melosh (1998) (cited in Collins, Melosh, & Marcus, 2005) obtained a method for estimating fine-grained ejecta sizes based on radar observations of parabola shaped features around craters on Venus. This characteristic distribution of fine ejecta is driven by zonal winds as they travel through the atmosphere. The finest particles are transported at furthest distances and take longer periods to settle according to Vervack & Melosh (1992) (cited in Collins, Melosh, & Marcus, 2005). Schaller & Melosh (1998) obtained an empirical law from comparing theoretical predictions of deposit formation and observations of the parabolic features that is used to estimate ejecta diameter $d_{ej}$ as shown below:

$$d_{ej} = d_c\left(\frac{D_{FR}}{2r_{km}}\right)^n \quad (41)$$

where the rim to rim diameter of the final crater $D_{fr}$ and the distance $r_{km}$ are in km. The value $d_c = 2400(D_{FR}/2)^{-1.62}$ while $n = 2.65$. However the relation applies for venusian zonal winds and atmospheric conditions and therefore is less reliable for Earth conditions especially in cases involving fine ejecta and fragmentability of sizable ejecta. According to Collins, Melosh, & Marcus (2005) this uncertainty is most pronounced within two transient crater radii from the crater center and at greater distances. Therefore in the present model, we calculate the ejecta thickness and diameters from the transient crater rim under the assumption that large-sized ejecta does not fragment upon landing.

### 2.3.4. Air Burst Effects

Large impacts produce shock waves within the atmosphere characterized by pressures greater than the normal atmospheric pressure (also known as peak overpressure) accompanied by winds travelling at high velocities. The nature of these effects are largely driven by impactor momentum and energy just after atmospheric entry. The assumption that large impactors achieve ground impacts while still relatively intact includes air burst events at the altitude $z = 0\ m$. The resulting effects of air burst are then determined in terms of Peak overpressure $P_o$, shock front velocity and peak wind velocity (Kring, 1997; Collins, Melosh, & Marcus, 2005). The estimation of these effects relies solely on yield scaling techniques performed on results of nuclear explosives detonation experiments on the ground surface.

Glasstone & Dolan (1977) (cited in Collins, Melosh, & Marcus, 2005) obtained empirical data on peak overpressure ($P_o$) in Pascals measured at different distances ($r_d$) in meters for a $1\ kt$ detonation on the ground surface. Collins, Melosh, & Marcus (2005) established a fit to this data and obtained the relation

$$P_o = \frac{P_c r_x}{4r_s}\left(1 + 3\left(\frac{r_x}{r_s}\right)^{1.3}\right) \quad (42)$$

where $P_c$ is the crossover pressure at distance $r_x = 290\ m$ where $P_o$ switches from $1/r^{2.3}$ behavior to $1/r$ behavior and has a value of $75000\ Pa$.

The overpressure generated by ground impacts are located within the Mach reflection region (Collins, Melosh, & Marcus, 2005). According to Glasstone & Dolan (1977) (cited in Collins, Melosh, & Marcus, 2005) this region enhances merging of pressure waves generated by impact which increase overpressures by factors that could reach up to two. The Equation 42 applies within this region which begins at the impact altitude $z = 0\ m$ while $r_x = 289\ m$ for all ground impacts (Collins, Melosh, & Marcus, 2005). Applied yield scaling demonstrates that "the ratio of distance at which a certain peak overpressure occurs to the cube root of the impact energy" is constant for impacts of any energy. Therefore the scaled distance rs for a $1\ kt$ impact that yields the relationship in Equation 42 is given by

$$r_s = \frac{r}{E_{kt}^{1/3}} \qquad (43)$$

Where $E_{kt}$ is the ground impact energy in kilotons.

According to Glasstone & Dolan (1977) (cited in Collins, Melosh, & Marcus, 2005), the maximum wind velocity $u$ is given by

$$u = \frac{5P_o}{7P_a} \frac{c_o}{(1 + 6P_o/7P_a)} \qquad (44)$$

where $P_a$ is the ambient pressure equal to $1\ bar$, $P_o$ is the peak overpressure and $c_o$ the ambient speed of sound assigned an approximate value $330\ m\ s^{-1}$. The time of arrival of the blast wave, $T_{bw}$, implemented in our model is given by

$$T_{bw} = \int_0^r \frac{dr}{U(r)} \qquad (45)$$

where the velocity of shock $U$ is given by

$$U(r) = c_o \left(1 + \frac{6P_o(r)}{7P_a}\right)^{1/2} \qquad (46)$$

Collins, Melosh, & Marcus (2005) suggest that the above model derived from small explosion experiments applies for atmospheric conditions characterized by uniform density. The resulting scaled distances do not account for the Earth's curvature in cases involving distances larger than $1\ km$ where peak overpressure might decay to zero. Therefore the effects of air burst by large impacts might not be reliably modelled since the scales of energies involved are significantly higher. Interaction between hot plumes of vaporized ejecta and the atmosphere are also not accounted for and might lead to overestimation of the air burst effects by factors ranging from 2 to 5.

## 3. Results and Discussion

The above review was used to develop a computational model with which simulations were performed for two near-Earth Asteroids. The first simulation yields a potential future impact event on Earth by object (29075) $1950DA$. The second simulation emulates a Chixculub scale hypothetical impact event by the Near-Earth Asteroid 433 $Eros$. Data on object (29075) 1950DA was obtained from ESA's Risk Page Special Risk List[2] . Data on Eros was obtained from Britt, Yeomans, Housen, & Consolmagno (2003) and Yeomans, et al. (2000) . The density $\rho_i$ for all simulated impacting objects was varied within the range $1000 - 8000\ kg\ m^{-3}$ for all possible impactor densities of asteroids (D.T Britt et al.). The entry angle $\theta$ between the impact trajectory and the horizontal ground surface was assumed to be $45°$ for typical entry events. The table below shows the various control variables as used in the model

| Object | $v_o\ (km\ s^{-1})$ | $L_o\ (m)$ | $\theta\ (°)$ | Reference |
|---|---|---|---|---|
| 29075 (1950DA) | 17.99 | 2000 | 45 | ESA Risk List |
| 433 $Eros$ | 20.00 | 17360 | 45 | Yeomans et al. (2000) |

Table 3: Input parameters for the three different simulations. Note that the fixed variables for each simulation are $v_i$, $L_i$ and $\theta$ while the manipulated variable is $\rho$

### 3.1. Atmospheric Entry Results

#### 3.1.1. Impactor Velocities

Figures 1 and 2 show the ratio by which the initial velocity $v_o$ is reduced for both objects varying with height $z$ . The velocity of each object decreases with increasing vertical distance from the Kármán line for any given initial density. The rate of change of velocity is highest for less dense impactors and lowest for more dense impactors. Object (29075) $1950DA$ is subjected to huge decrease in velocity than Eros upon ground impact occurring at distance $100000\ m$. The ratio of impact velocity $v_f$ to the entry velocity $v_o$ ranges from 0.99099 to 0.99893 for $1950DA$ while that of $Eros$ ranges from 0.99902 to 0.99988 for all impactor densities.

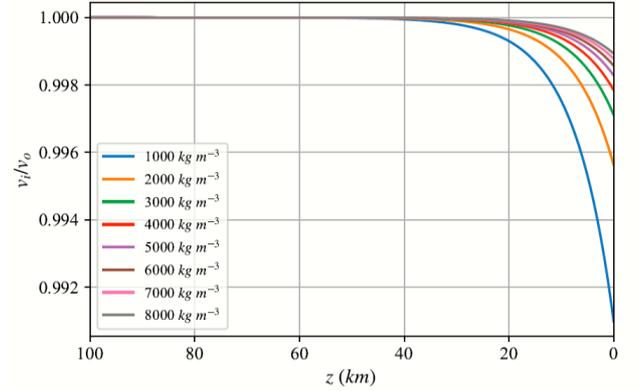

Figure 1: The ratio by which impactor velocities change $v_i/v_o$ as a function of height $z$ from the ground to the Kármán Line for Object (29075) $1950DA$ assigned different initial densities

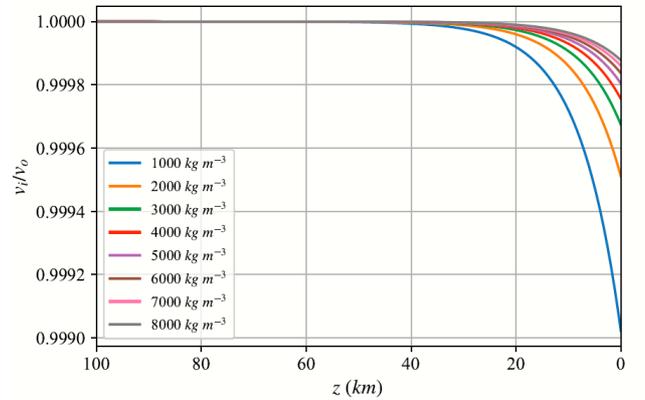

Figure 2: The ratio by which impactor velocities change $v_i/v_o$ as a function of height $z$ from the ground to the Kármán Line for Object 433 $Eros$ assigned different initial densities

#### 3.1.2. Stagnation Pressure

Figures 3 and 4 show the value of the ratio of stagnation pressure $P_{si}$ to the initial entry pressure $P_{so}$ acting on each object varying with height $z$ from the ground surface. For any given initial density, the ratio value $P_{si}/P_{so}$ of pressures acting on each object at the same distance $z$ is relatively similar. Both objects experience increasing pressure $P_{si}$ as they move towards the ground where $z = 100000\ m$.

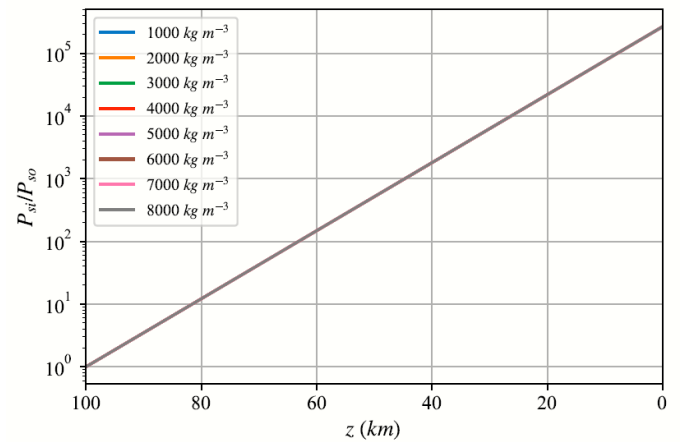

Figure 3: The log-linear plot of ratios by which stagnation pressures change $P_{si}/P_{so}$ as a function of height $z$ from the ground to the Kármán Line for Object (29075) $1950DA$ assigned different initial densities

---

[2] Data on objects identified in the Risk List can be changed or even removed based on observation updates. The reference was made on 6th December 2019

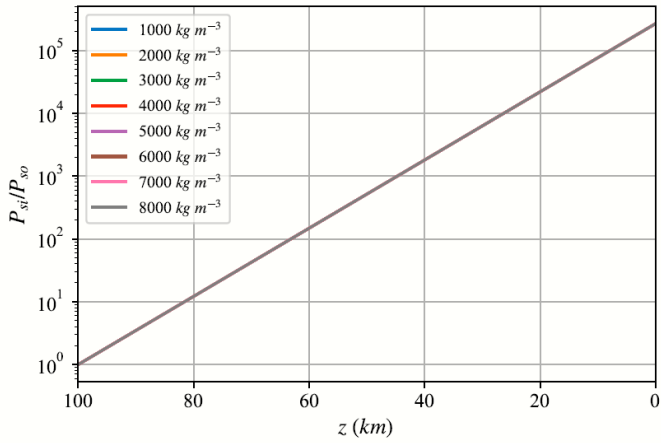

Figure 4: The log-linear plot of ratios by which stagnation pressures change $P_{si}/P_{so}$ as a function of height $z$ from the ground to the Kármán Line for Object 433 $Eros$ assigned different initial densities

### 3.1.3. Impactor Diameters

Figures 5 and 6 show the value of the ratio of impactor diameter $L_i$ to the initial diameter $L_o$ varying with height $z$ for each object. Both objects show increasing diameters as they move towards the ground for any given initial density. In both cases, rate of change of diameter $L_i$ is highest for impactors with low initial density and lowest for impactors with high intial density. Object (29075) $1950DA$ hits the ground with diameter ratio values ranging from 1.0080 to 1.06611 while that of Object 433 $Eros$ ranges from 1.00011 to 1.00085

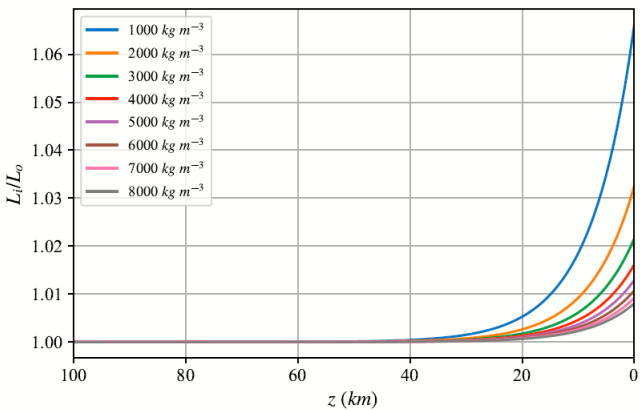

Figure 5: The ratio by which impactor diameters change $L_i/L_o$ as a function of height $z$ from the ground to the Kármán Line for Object (29075) $1950DA$ assigned different initial densities

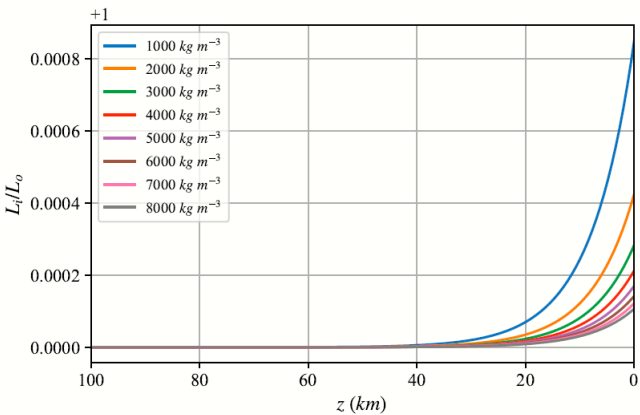

Figure 6: The ratio by which impactor diameters change $L_i/L_o$ as a function of height $z$ from the ground to the Kármán Line for Object 433 $Eros$ assigned different initial densities

### 3.1.4. Impactor Densities

Figures 7 and 8 show fractional value by which the impactor density $\rho_o$ is reduced varying with distance $D$ for both objects. The density $\rho_i$ of each object decreases with decreasing height from the ground surface. In both cases, the rate of change in density is highest for objects with low initial density and lowest for impactors with high initial densities. Upon ground impact, object (29075) $1950DA$ has ratio values ranging from 0.82527 to 0.97628 while those of Eros range from 0.99745 to 0.99968.

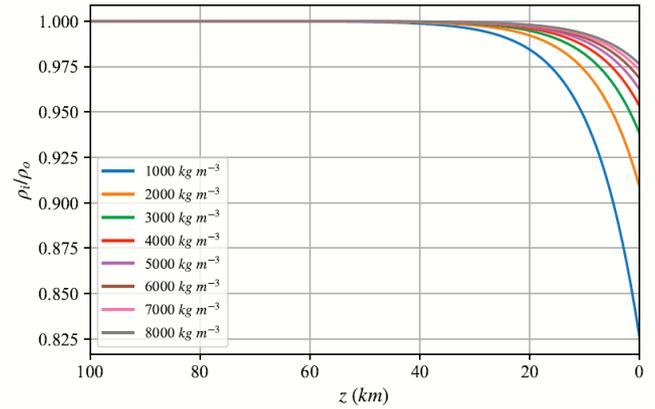

Figure 7: The ratio by which initial impactor density changes $\rho_i/\rho_o$ as a function of height $z$ from the ground to the Kármán Line for Object (29075) $1950DA$ assigned different initial densities

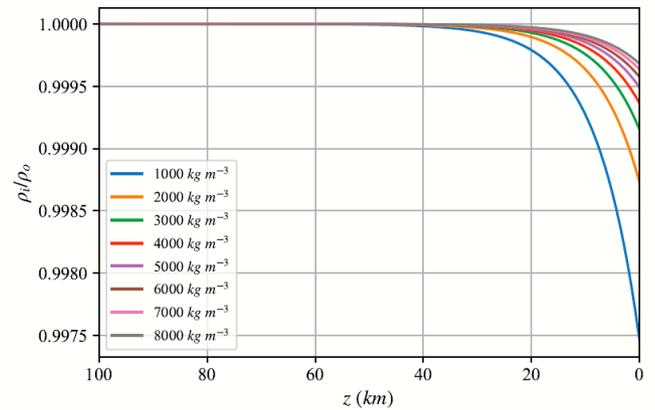

Figure 8: The ratio by which initial impactot density changes $\rho_i/\rho_o$ as a function of height $z$ from the ground to the Kármán Line for Object 433 $Eros$ assigned different initial densities

### 3.1.5. Impcator Yield Strengths

Figures 9 and 10 show value of the ratio of impactor strength $Y_i$ to the initial strength $Y_o$ for each object varying with decreasing height $z$. The strength of each object $Y_i$ decreases with increasing vertical distance from the Kármán Line. This rate of change in impactor strength is highest for objects with low initial density and lowest for objects with low initial density. Object (29075) $1950DA$ has ratio values ranging from 0.65969 to 0.85785 upon ground impact while those of Eros are in the range 0.99423 to 0.99794.

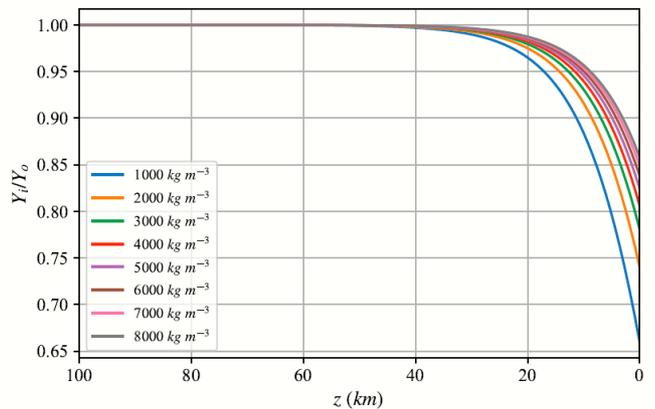

Figure 9: The ratio by which impactor Yield Strength changes $Y_i/Y_o$ as a function of height $z$ from the ground to the Kármán Line for Object (29075) $1950DA$ assigned different initial densities

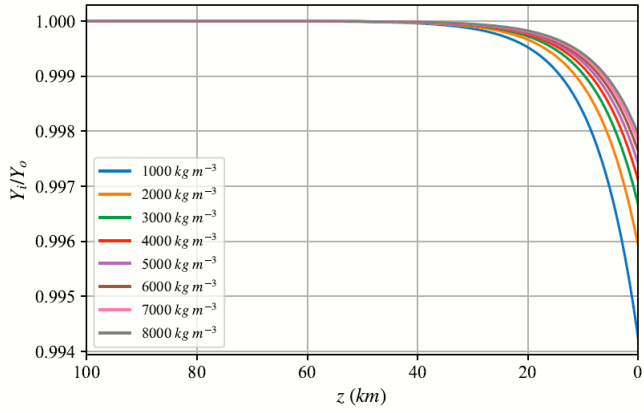

Figure 10: The ratio by which impactor Yield Strength changes $Y_i/Y_o$ as a function of height $z$ from the ground to the Kármán Line for Object 433 $Eros$ assigned different initial densities

### 3.1.6. Impactor Energies

Figures 11 and 12 show the value of the ratio of impactor energy $E_i$ to the initial entry energy $E_o$ for each object decreasing with height $z$. The energies of both objects decrease as they move towards the ground. Objects with low initial densities demonstrate a higher rate of change in energy while those with low initial densities have a lower rates of change in energy. Object (29075) 1950$DA$ has final ratios ranging from 0.98206 to 0.99786 upon ground impact while Eros has values ranging from 0.99804 to 0.99976.

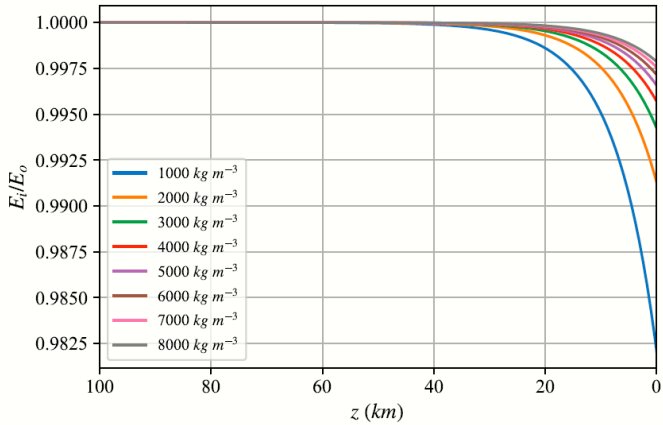

Figure 11: The ratio by which initial impactor energy changes $E_i/E_o$ as a function of height $z$ from the ground to the Kármán Line for Object (29075) 1950$DA$ assigned different initial densities

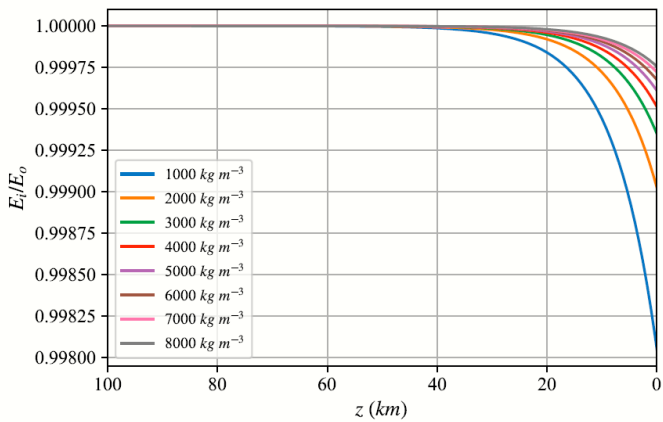

Figure 12: The ratio by which initial impactor energy changes $E_i/E_o$ as a function of height $z$ from the ground to the Kármán Line for Object 433 $Eros$ assigned different initial densities

Figures 13 and 14 show value of the ratio of final impactor densities $\rho_f$ to the initial impactor densities $\rho_o$ plotted against initial densities for both objects. On impact, objects with the least initial densities show huge decrease in their density while those with high initial densities show small decrease in density. The rate of change in density decreases as initial densities approach 8000 $kg\ m^{-3}$. Object (29075) 1950$DA$ experiences the highest change in initial density with ratio values ranging from 0.825 to 0.976 upon ground impact. Object 433 $Eros$ experiences lowest change in density with ratios ranging from 0.9974 to 0.9997 upon completimg atmospheric entry.

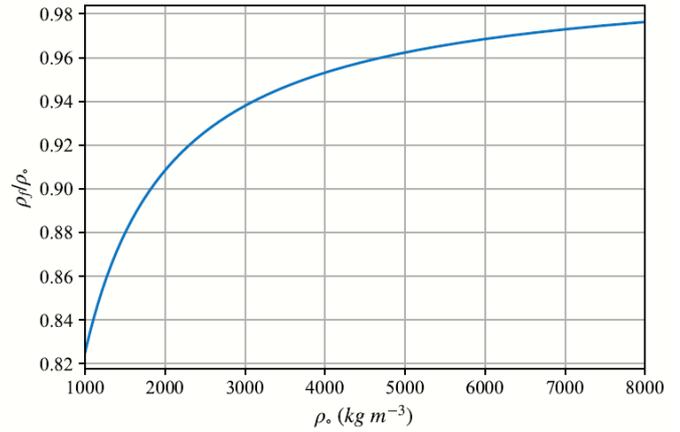

Figure 13: The ratio value by which impactor density changes $\rho_f/\rho_o$ as a function of height $z$ from the ground to the Kármán Line for Object (29075) 1950$DA$ assigned different initial densities $\rho_o$.

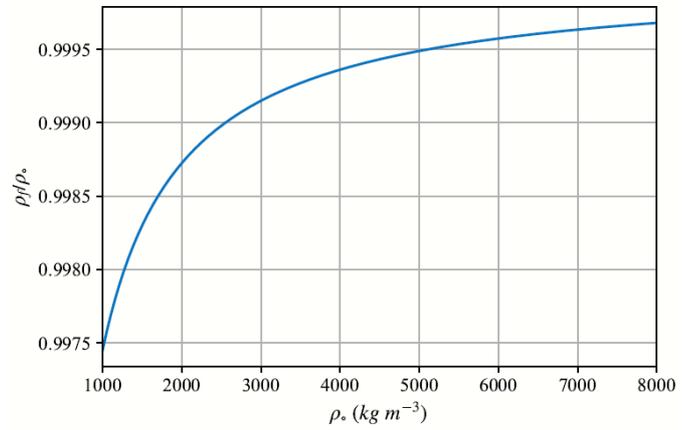

Figure 14: The ratio value by which impactor density changes $\rho_f/\rho_o$ as a function of height $z$ from the ground to the Kármán Line for Object 433 $Eros$ assigned different initial densities $\rho_o$.

Tables 4 and 5 show the final parameters from atmospheric entry just before ground impact. Both objects show decreasing altitudes of breakup $z_{br}$ and final diameters $L_f$ with increasing initial densities. For fixed initial densities, Object 433 $Eros$ has slightly higher altitudes of breakup and larger final diameters than Object (29075) 1950$DA$. Increasing intial densities for both objects show increasing final velocity $v_f$, final energy $E_f$ and final density $\rho_f$ . However Object 433 $Eros$ has higher values for final velocity, final energy and final density for any given initial density.

| $\rho_o$ (kg m$^{-3}$) | $z_{br}$ (m) | $v_f$ (m s$^{-1}$) | $L_f$ (m) | $E_f$ (J) | $\rho_f$ (kg m$^{-3}$) |
|---|---|---|---|---|---|
| 1000 | 8.1600e+4 | 17827.8827 | 2132.2151 | 6.6567e+20 | 825.2717 |
| 2000 | 6.6544e+4 | 17911.3509 | 2065.0411 | 1.3438e+21 | 1816.9121 |
| 3000 | 5.4993e+4 | 17938.0895 | 2043.1292 | 2.0218e+21 | 2813.9976 |
| 4000 | 4.5257e+4 | 17951.2611 | 2032.2607 | 2.6997e+21 | 3812.5165 |
| 5000 | 3.6686e+4 | 17959.1014 | 2025.7681 | 3.3775e+21 | 4811.6140 |
| 6000 | 2.8950e+4 | 17964.3024 | 2021.4510 | 4.0554e+21 | 5811.0091 |
| 7000 | 2.1863e+4 | 17968.0047 | 2018.3730 | 4.7332e+21 | 6810.5740 |
| 8000 | 1.5318e+4 | 17970.7746 | 2016.0677 | 5.4111e+21 | 7810.2448 |

Table 4: Altitudes of breakup ($z_{br}$), final velocities ($v_f$), final diameters ($L_f$), final energies ($E_f$) and final densities ($\rho_f$) for Object (29075) 1950DA assigned different initial densities $\rho_o$

| $\rho_o$ (kg m$^{-3}$) | $z_{br}$ (m) | $v_f$ (m s$^{-1}$) | $L_f$ (m) | $E_f$ (J) | $\rho_f$ (kg m$^{-3}$) |
|---|---|---|---|---|---|
| 1000 | 8.3295e+4 | 19980.4382 | 17347.7662 | 5.4680e+23 | 997.4526 |
| 2000 | 6.8238e+4 | 19990.2195 | 17367.3868 | 1.0947e+24 | 1997.4491 |
| 3000 | 5.6685e+4 | 19993.4792 | 17364.9271 | 1.6425e+24 | 2997.4471 |
| 4000 | 4.6946e+4 | 19995.1085 | 17363.7011 | 2.1904e+24 | 3997.4427 |
| 5000 | 3.8365e+4 | 19996.0862 | 17362.9630 | 2.7383e+24 | 4997.4407 |
| 6000 | 3.0608e+4 | 19996.7380 | 17362.4709 | 3.2861e+24 | 5997.4388 |
| 7000 | 2.3474e+4 | 19997.2036 | 17362.1194 | 3.8340e+24 | 6997.4369 |
| 8000 | 1.6835e+4 | 19997.5528 | 17361.8556 | 4.3819e+24 | 7997.4352 |

Table 5: Altitudes of breakup ($z_{br}$), final velocities ($v_f$), final diameters ($L_f$), final energies ($E_f$) and final densities ($\rho_f$) for Object 433 Eros assigned different initial densities $\rho_o$

| | $a$ | $b$ | $R^2$ | $S$ (kg m$^{-3}$) |
|---|---|---|---|---|
| Fig. 15 | −180.1541 ± 0.1395 | 0.9985 ± 2.83e − 05 | 99% | 1.6194 |
| Fig. 16 | −2.5462 ± 5.67e − 05 | 0.999998 ± 1.15e − 08 | 99% | 1.6194 |

Table 6: Values of parameters $a$ and $b$ for linear fit models in Figs. 16 and 15

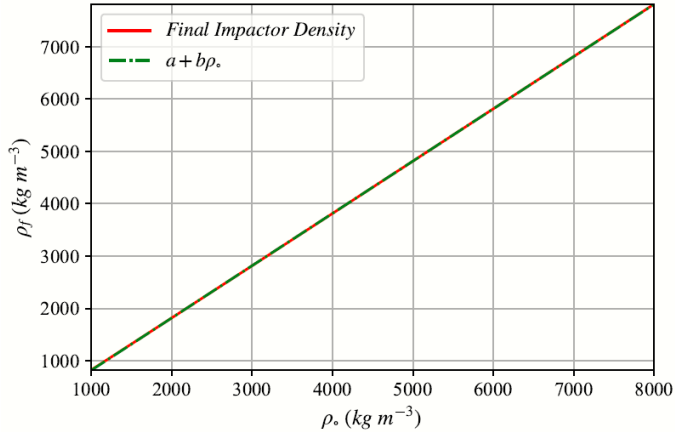

Figure 15: The final density $\rho_i$ against different initial density $\rho_o$ for Object (29075) 1950DA fitted with a linear model

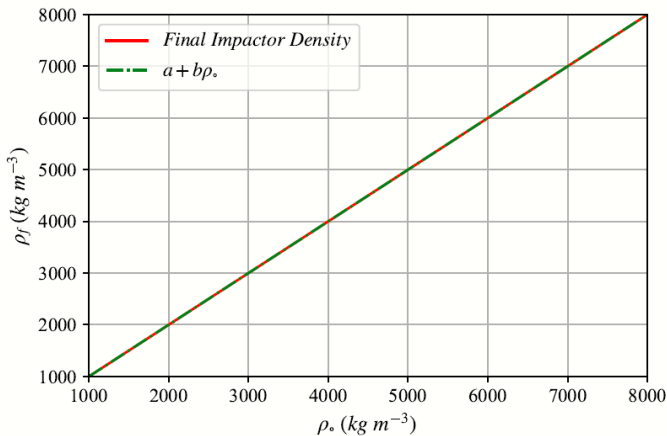

Figure 16: The final density $\rho_i$ against different initial density $\rho_o$ for Object 433 Eros fitted with a linear model

Figures 15 and 16 show linear fits for final density as a function of initial density. Parameters a and b are given in the table below

Given similar initial densities, objects 1950DA and Eros have the same strength characteristics prior to entry but different entry velocities and diameters. The larger diameter and slightly higher velocity by Eros means it is subjected to more stagnation pressure upon entry than 1950DA. Tables 4 and 5 show that Eros breaks up at a slightly higher altitude than 1950DA given the same initial density and strength. However, Figures 5 and 6 show that the rate of expansion of the initial diameter is higher for 1950DA than Eros. An inference is made here that the difference in mass of the two objects determines the amount by which the initial diameters are increased. Given fixed initial densities, Eros contains more mass than 1950DA. Therefore forces acting horizontal to the direction of motion are not sufficient enough to expand Eros' massive fragments at a rate faster than that of 1950DA.

1950DA has a slightly lower entry velocity than Eros but a much smaller diameter. Deductions from the above inference imply that 1950DA expands at a much faster rate than Eros. Therefore the rate at which the atmosphere is intercepted after breakup by 1950DA is higher than that of Eros. This interception in turn reduces the initial velocity of 1950DA at a much faster rate than Eros as shown in Figures 1 and 2. An inference is made that the amount by which the initial velocity is reduced depends on the rate of diameter expansion.

Throughout the entry process, the densities of each object is dependent on its respective diameter while its mass remains constant. Object 1950DA has higher diameter expansion rate and therefore a higher rate of change in density than Eros. This effect on density is as shown in Figures 7 and 8. Similarly, strength of each object, also dependent on object density, show faster rate of decrease for 1950DA than Eros. This is as shown in Figures 9 and 10.

For each of the above observed parameters, the curve shapes for both objects given similar initial densities are similar. Figures 3 and 4 show that the initial stagnation pressures are increased by similar fractional values for equal distances traversed by each object given similar initial densities. This implies that the nature of effect from atmospheric disruption is prevalent in both cases. This atmospheric effect is due to the exponential nature of stagnation pressure acting on the objects in the entry process. Therefore the exponential effect of the Earth's atmosphere an all entry parameters means that these effects are maximized just before ground impact. For each individual object, increasing the initial object density results to decreasing difference between consecutive parameter values at any given distances from the top of the atmosphere. This difference between consecutive values also increases with increasing distance from the top of the atmosphere.

3.2. Crater Formation and Collapse

Tables 7 and 8 show results from crater formation events for object $1950DA$ and $Eros$ respectively. For fixed initial densities, craters formed by Eros are wider and deeper than those of $1950DA$. $Eros$ also melts greater quantities melt than $1950DA$, resulting in thicker melt sheets at the crater base than $1950DA$. Each object shows increasing crater depth, diameter, melt volume and thickness with increasing initial density.

| $\rho_o$ (kg m$^{-3}$) | $D_{FR}$ (m) | $d_{FC}$ (m) | $V_M$ (m$^{-3}$) | $t_M$ (m) |
|---|---|---|---|---|
| 1000 | 16856.2 | 688.0 | 5.0411e+09 | 35.3 |
| 2000 | 21431.5 | 739.6 | 1.0177e+10 | 44.1 |
| 3000 | 24606.8 | 771.0 | 1.5311e+10 | 50.3 |
| 4000 | 27123.8 | 793.9 | 2.0445e+10 | 55.3 |
| 5000 | 29244.4 | 812.1 | 2.5578e+10 | 59.5 |
| 6000 | 31095.4 | 827.3 | 3.0711e+10 | 63.2 |
| 7000 | 32748.9 | 840.3 | 3.5845e+10 | 66.5 |
| 8000 | 34250.4 | 851.7 | 4.0978e+10 | 69.5 |

Table 7: Final crater diameters ($D_{FR}$), final crater depth ($d_{FC}$), melt volume ($V_M$) and thickness of melt ($t_M$) values produced by Object (29075) 1950DA assigned different initial densities ro after hitting a target with density $\rho_t = 2500\ kg\ m^{-3}$ (sedimentary)

| $\rho_o$ (kg m$^{-3}$) | $D_{FR}$ (m) | $d_{FC}$ (m) | $V_M$ (m$^{-3}$) | $t_M$ (m) |
|---|---|---|---|---|
| 1000 | 96966.5 | 1165.0 | 4.1409e+12 | 876.2 |
| 2000 | 122207.8 | 1249.0 | 8.2900e+12 | 1104.3 |
| 3000 | 139907.4 | 1300.9 | 1.2439e+13 | 1264.3 |
| 4000 | 153996.0 | 1339.0 | 1.6588e+13 | 1391.6 |
| 5000 | 165892.2 | 1369.4 | 2.0737e+13 | 1499.1 |
| 6000 | 176290.4 | 1394.6 | 2.4886e+13 | 1593.1 |
| 7000 | 185588.4 | 1416.4 | 2.9035e+13 | 1677.1 |
| 8000 | 194037.7 | 1435.5 | 3.3184e+13 | 1753.4 |

Table 8: Final crater diameters ($D_{FR}$), final crater depth ($d_{FC}$), melt volume ($V_M$) and thickness of melt ($t_M$) values produced by Object 433 $Eros$ assigned different initial densities ro after hitting a target with density $\rho_t = 2500\ kg\ m^{-3}$ (sedimentary)

Figure 17 and 18 show final crater diameters plotted against initial object densities for objects 1950DA and Eros respectively. Both figures show that the difference between consecutive crater diameter values decreases with increasing initial densities. In each case, the resulting final crater diameter obeys a power law relationship whose parameter values are as shown in Table 9 below. These values are determined by the distinct effects atmospheric entry has on each object. $Eros$ demonstrates Chixculub scale crater diameters for densities greater than approximately 6000 $kg\ m^{-3}$.

| | $a$ | $b$ | $S$ (m) |
|---|---|---|---|
| Fig. 17 | $1631.02 \pm 0.81$ | $0.3389 \pm 5.86e-05$ | 21.44 |
| Fig. 18 | $9685.65 \pm 0.16$ | $0.3335 \pm 1.99e-06$ | 4.15 |

Table 9: Values of parameters $a$ and $b$ for non-linear fit models in Figs. 18 and 17

Cratering processes are largely dependent on final impactor diameter $v_f$, density $\rho_f$ and diameter $L_f$. These values represent the energy of the collision of the impactor with the ground after atmospheric disruption. $Eros$ experiences much less disruption during entry than 1950DA and therefore impacts the ground with greater energy hence displaces and melts greater quantities of material. In the case of each object, increasing the initial density translates into increasing energy of collision with the ground. Thus increasingly greater quantities are excavated and melted when the initial density of a single object is increased.

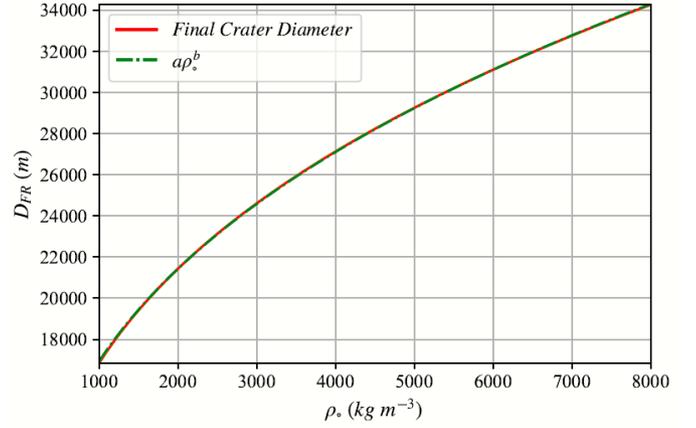

Figure 17: The final crater diameter $D_{FC}$ against initial impactor density $r_o$ for Object (29075) 1950DA fitted with a non-linear model

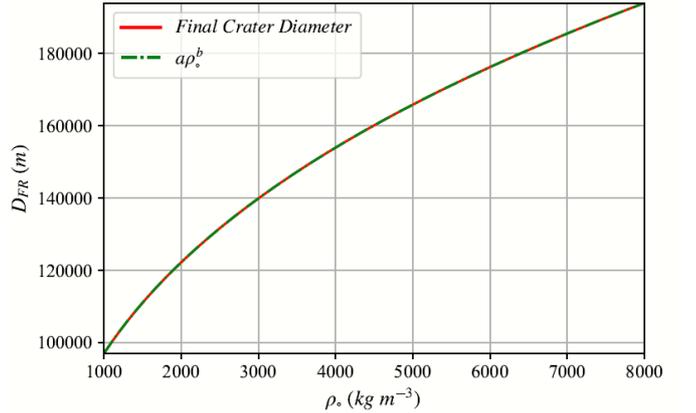

Figure 18: The final crater diameter $D_{FC}$ against initial impactor density $r_o$ for Object 433 $Eros$ fitted with a non-linear model

3.3. Thermal Radiation

Tables 10 and 11 show fireball radii lengths, thermal exposure at fireball radii, time at which radiation is maximum and duration of irradiation of the fireball for objects (29075) 1950DA and $Eros$ respectively. At similar initial densities, impacts by $Eros$ generate fireballs with greater radii than 1950DA. Fireballs generated by Eros also take longer periods for radiation to escape and be completely irradiated than those of 1950DA.

| $\rho_o$ (kg m$^{-3}$) | $R_{fb}$ (m) | $j\alpha R_{fb}$ (J m$^{-2}$) | $T_{max}$ (s) | $T_i$ (s) |
|---|---|---|---|---|
| 1000 | 17462.9 | 1.0404e+09 | 0.98 | 226.93 |
| 2000 | 22070.5 | 1.3143e+09 | 1.23 | 286.81 |
| 3000 | 25289.5 | 1.5055e+09 | 1.41 | 328.64 |
| 4000 | 27848.4 | 1.6575e+09 | 1.55 | 361.90 |
| 5000 | 30007.5 | 1.7856e+09 | 1.67 | 389.95 |
| 6000 | 31893.8 | 1.8975e+09 | 1.78 | 414.47 |
| 7000 | 33580.1 | 1.9974e+09 | 1.87 | 436.38 |
| 8000 | 35112.1 | 2.0882e+09 | 1.95 | 456.29 |

Table 10: Fireball radius $R_{fb}$, corrected thermal exposure at fireball radius $j\alpha R_{fb}$, time of maximum radiation $T_{max}$ and duration of irradiation $T_i$ values from impact by Object (29075) 1950DA assigned different initial densities $\rho_o$

Figures 19 and 20 show the corrected thermal exposure values at varying distances from the site of impact for $Eros$ and 1950DA respectively. Exposure to radiation begins at fireball radii after radiation escapes the respective fireballs.

| $\rho_o$ (kg m$^{-3}$) | $R_{fb}$ (m) | $j\alpha R_{fb}$ (J m$^{-2}$) | $T_{max}$ (s) | $T_i$ (s) |
|---|---|---|---|---|
| 1000 | 163545.8 | 9.6014e+09 | 8.19 | 2125.31 |
| 2000 | 206122.0 | 1.2049e+10 | 10.31 | 2678.60 |
| 3000 | 235976.4 | 1.3752e+10 | 11.80 | 3066.56 |
| 4000 | 259739.8 | 1.5100e+10 | 12.99 | 3375.37 |
| 5000 | 279805.3 | 1.6232e+10 | 13.99 | 3636.13 |
| 6000 | 297344.0 | 1.7219e+10 | 14.87 | 3864.05 |
| 7000 | 313026.7 | 1.8098e+10 | 15.65 | 4067.85 |
| 8000 | 327278.3 | 1.8894e+10 | 16.37 | 4253.05 |

Table 11: Fireball radius $R_{fb}$, corrected thermal exposure at fireball radius $j\alpha R_{fb}$, time of maximum radiation $T_{max}$ and duration of irradiation $T_i$ values from impact by Object 433 $Eros$ assigned different initial densities $\rho_o$

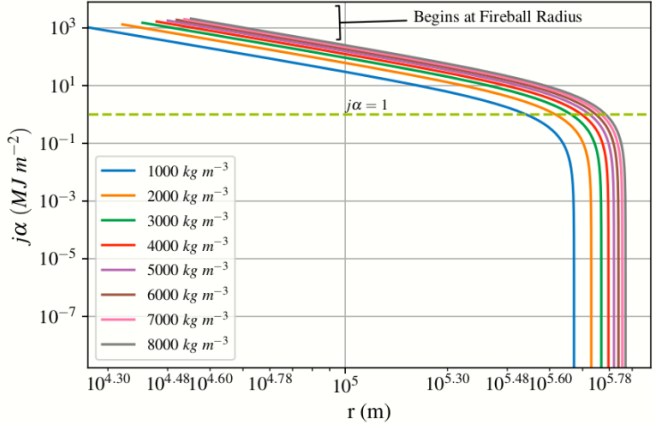

Figure 19: Thermal exposure $j\alpha$ plotted against distance r for Object (29075) 1950$DA$ assigned different initial densities

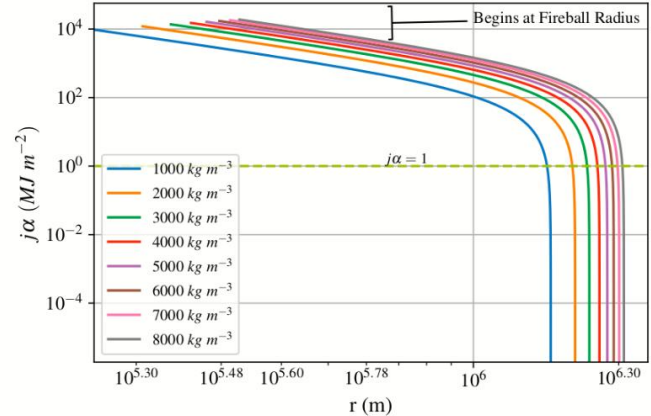

Figure 20: Thermal exposure $j\alpha$ plotted against distance r for Object 433 $Eros$ assigned different initial densities

Given identical initial densities, Eros has greater thermal exposure values than 1950$DA$ at similar distances. However, thermal exposure by 1950$DA$ decreases at faster rates and reaches zero at closer distances than exposure by $Eros$. To allow for determination of ignited materials given in Table 1, thermal exposure values are given in Mega Joules $MJ$. Line $j\alpha$ marks distance within which all mentioned materials are ignited and all degrees of burns sustained.

In the case of each object, increasing the initial density shows increasing fireball radii, times of maximum radiation and irradiation durations. At similar distances, less dense objects yield lower thermal exposure values while more dense objects generate higher thermal exposure values. However there difference between these consecutive values decrease with increasing initial density. Exposure to radiation however decreases at faster rates for objects with low initial densities than those with high initial densities.

Eros maintains greater amounts of momentum even after atmospheric disruption compared to 1950$DA$. Therefore the fraction of ground impact that generates thermal radiation yields greater thermal energy for $Eros$ than 1950$DA$. In the case of each object, the thermal energy generated increases with when initial density is increased. Figures 21 and 22 show increasing fireball radii values with increasing density, demonstrating a power law relationship. Both figures show that the difference between consecutive fireball radii decreases with increasing initial densities. The parameter values, shown in Table 12 below are determined by the distinct entry and impact characteristics of each object.

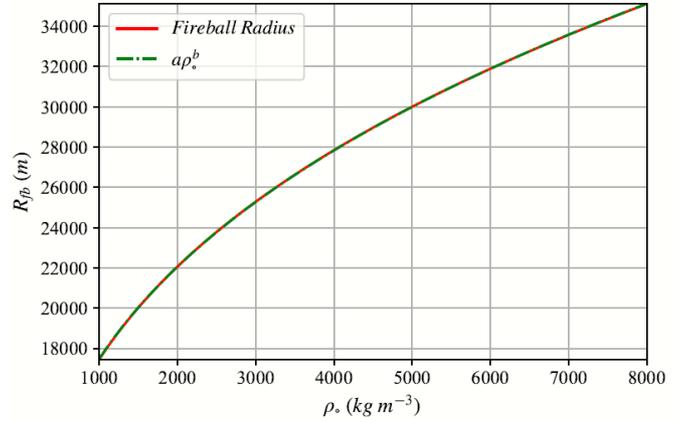

Figure 21: The fireball radius $R_{fb}$ against initial impactor density $\rho_o$ for Object (29075) 1950$DA$ fitted with a non-linear model

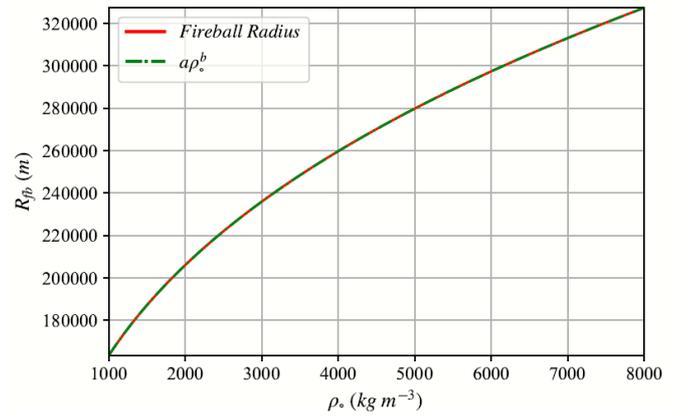

Figure 22: The fireball radius $R_{fb}$ against initial impactor density $\rho_o$ for Object 433 $Eros$ fitted with a non-linear model

|  | $a$ | $b$ | $S$ (m) |
|---|---|---|---|
| Fig. 21 | $1727.56 \pm 0.29$ | $0.34 \pm 2.02e-05$ | 7.58 |
| Fig. 22 | $16334.98 \pm 0.29$ | $0.33 \pm 2.10e-06$ | 7.39 |

Table 12: Values of parameters $a$ and $b$ for non-linear fit models in Figs. 22 and 21

3.4. Seismic Effects

Figures 23 and 24 show the effective magnitude at various distances from impact site for objects (29075)1950$DA$ and 430 $Eros$ respectively. In both cases, the effective magnitude decreases with increasing distance from the site of impact. Given similar initial densities, the magnitude of seismic waves is greater for $Eros$ than 1950$DA$ at similar distances. For individual objects, lower initial densities yield seismic waves weaker in magnitude than those yielded by higher initial densities at similar distances. The effective magnitude decreases with increasing distance at similar rates upto distances close to 60000 $m$ where more dense impactors have higher rates of decrease than less dense impactors. Similar effects are observed for distances beyond 60000 $m$ until distances close to 700000 $m$. Beyond 700000 $m$, the rate of decrease of seismic magnitude is similar for all densities.

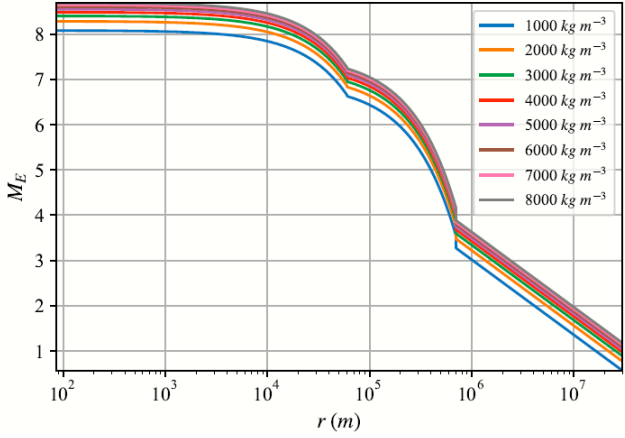

Figure 23: Effective magnitude $M_e$ plotted against distance from the epicenter r for Object (29075) 1950DA assigned different initial densities

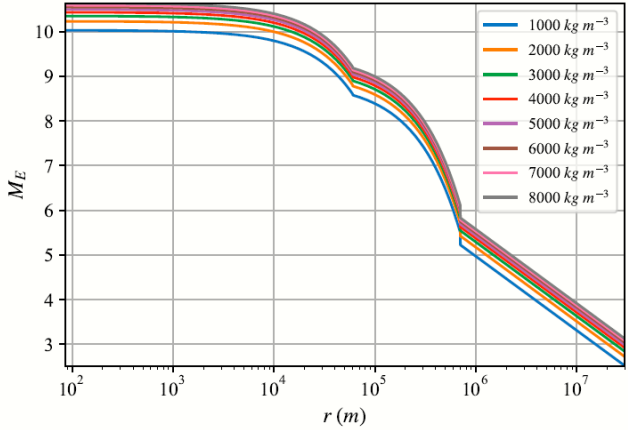

Figure 24: Effective magnitude $M_e$ plotted against distance from the epicenter r for Object 430 $Eros$ assigned different initial densities

Figures 25 and 26 show the time of arrival of the most energetic waves at different distances from the impact site for objects 1950DA and $Eros$ respectively. Both objects show similar arrival times at similar distances.

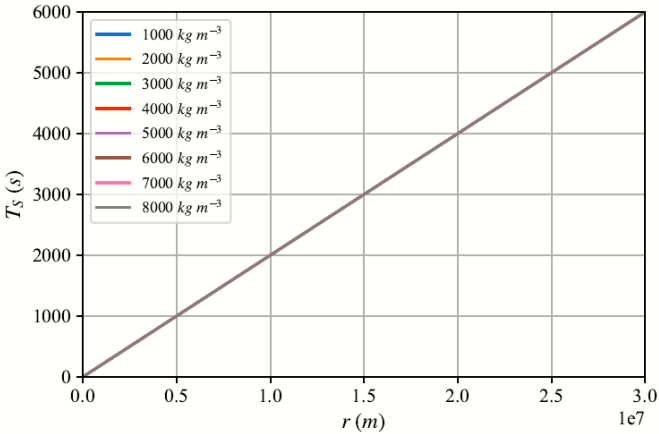

Figure 25: Seismic arrival time $T_s$ plotted against distance from the epicenter for Object (29075) 1950DA assigned different initial densities

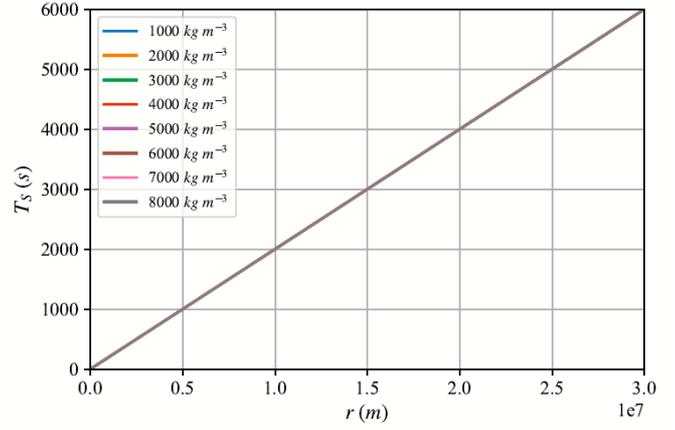

Figure 26: Seismic arrival time $T_s$ plotted against distance from the epicenter for Object 433 $Eros$ assigned different initial densities

Figures 27 and 28 show the seismic magnitudes plotted against different initial densities for objects 1950DA and $Eros$ respectively. The difference between consecutive seismic magnitudes decreases with increasing initial object densities. The resulting curves are described by a linear model whose parameters are shown in Table 13 below. The parameter values are specific for each object given distinct initial properties and entry effects.

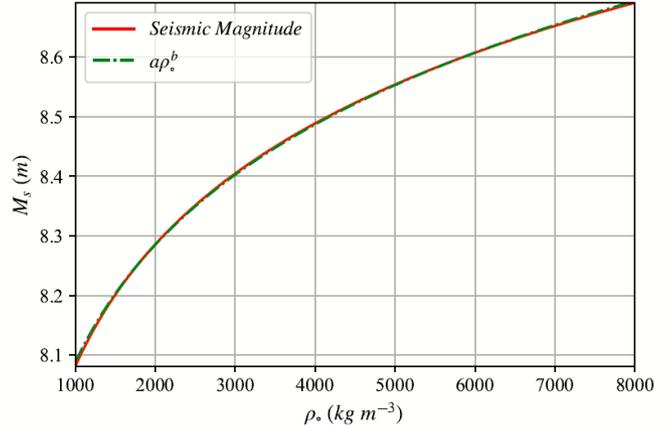

Figure 27: Seismic Magnitude $M_s$ plotted for different initial impactor densities $\rho_o$ for Object 29075 (1950DA) fitted with a non-linear model

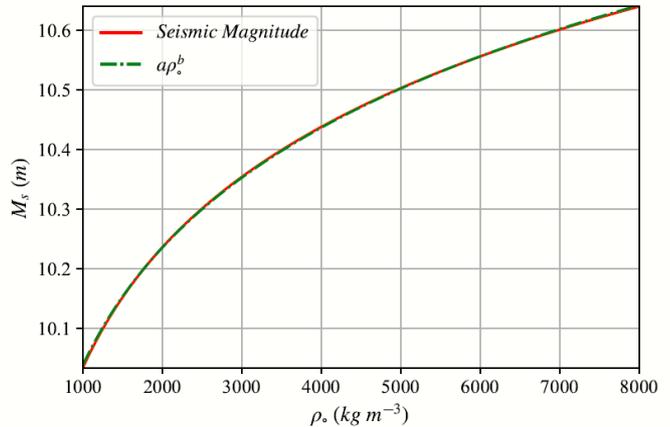

Figure 28: Seismic Magnitude $M_s$ plotted for different initial impactor densities $\rho_o$ for Object 433 $Eros$ fitted with a non-linear model

|  | $a$ | $b$ | $S\ (m)$ |
| --- | --- | --- | --- |
| Fig. 27 | $6.36 \pm 7.34e - 04$ | $0.03 \pm 1.39e - 05$ | 0.00 |
| Fig. 28 | $8.27 \pm 5.4e - 04$ | $0.03 \pm 7.79e - 06$ | 0.00 |

Table 13: Values of parameters $a$ and $b$ for non-linear fit models in Figs. 28 and 27

The fraction of impact energy that generates seismic effects yields greater effects for Eros than 1950DA. Therefore $Eros$ has greater momentum prior to ground impact than 1950$DA$. This supports the fact that Eros experiences less disruption during atmospheric entry than 1950$DA$. For a single object, increasing the initial density results in greater momentum and therefore greater seismic energy.

### 3.5. Ejecta Deposition

Figures 29 and 30 show ejecta thickness at various distances from the crater given different initial densities for object 1950$DA$ and $Eros$ respectively. For similar initial densities, Eros shows thicker ejecta properties than 1950$DA$. For each object, increasing the initial object density shows increasing ejecta thickness at similar distances. The difference between consecutive ejecta thickness values also decreases with increasing initial object density. For each object, the ejecta thickness decreases with increasing distance at similar rates for all initial densities.

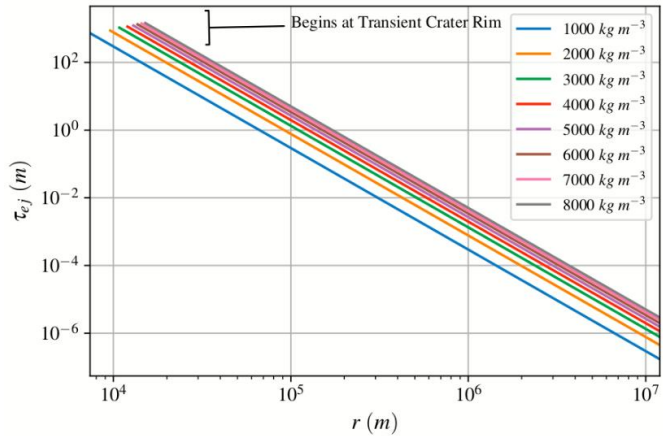

Figure 29: Ejecta thickness $\tau_{ej}$ against distance r plotted for Object (29075) 1950$DA$ assigned different initial densities

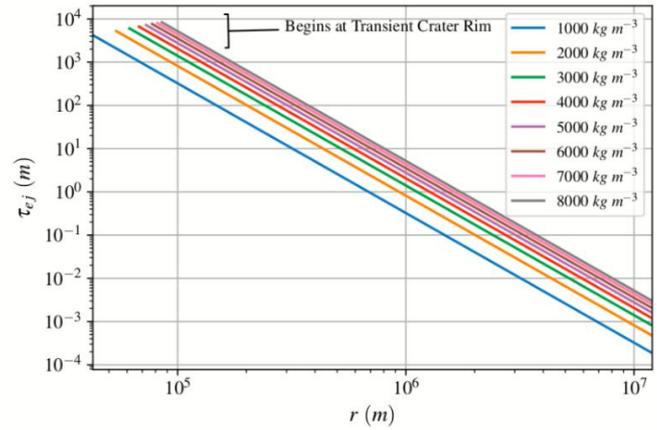

Figure 30: Ejecta thickness $\tau_{ej}$ against distance r plotted for Object 433 $Eros$ assigned different initial densities

Figure 31 and 32 show decreasing ejecta diameters with increasing distance from the transient crater rim for 1950$DA$ and $Eros$ respectively. Given similar initial densities, $Eros$ yields ejecta sizes larger than those of 1950$DA$ at similar distances. For each object, lower initial densities yield lower ejecta sizes while higher initial densities yield large sized ejecta at similar distances. The difference between consecutive ejecta values also decreases with increasing initial object density at fixed distances.

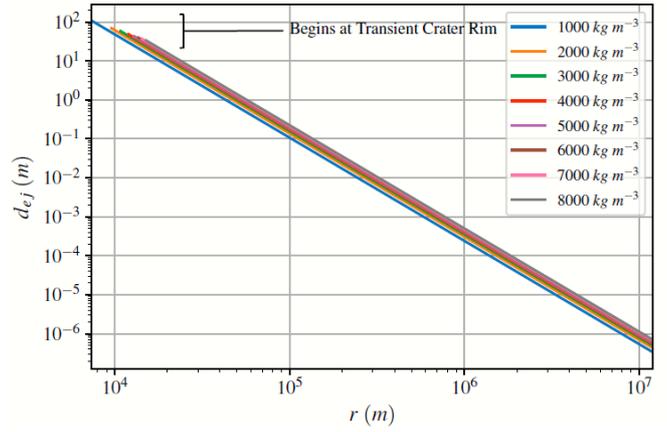

Figure 31: Ejecta diameter $d_{ej}$ against distance $r$ plotted for Object (29075) 1950$DA$ assigned different initial densities

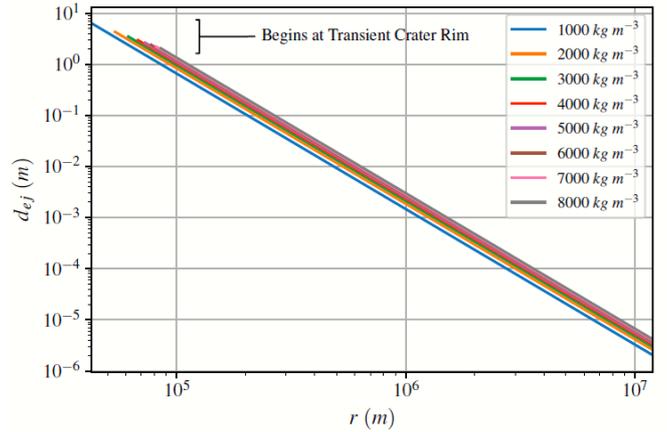

Figure 32: Ejecta diameter $d_{ej}$ against distance $r$ plotted for Object 433 $Eros$ assigned different initial densities

Figures 33 and 34 show increasing Transient crater volumes and Ejecta volumes with increasing initial density values for objects 1950$DA$ and $Eros$ respectively. For fixed initial densities, Eros excavates greater quantities of ejecta material than 1950$DA$.

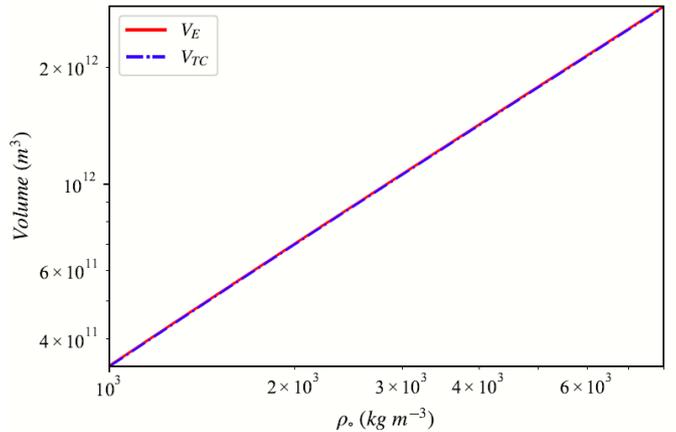

Figure 33: Ejecta Volume $V_E$ and Transient Crater Volume $V_{TC}$ plotted against initial density $\rho_o$ for Object (29075) 1950$DA$ assigned different initial densities

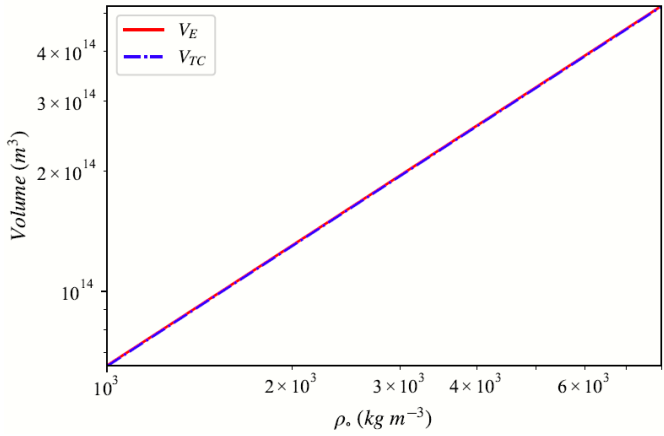

Figure 34: Ejecta Volume $V_E$ and Transient Crater Volume $V_{TC}$ plotted against initial density $\rho_o$ for Object 433 *Eros* assigned different initial densities

Ejecta deposition effects are driven by crater excavation and displacement of ground material by objects in a ground impact event. *Eros* has greater momentum upon collision with the ground since it is less disrupted by atmospheric entry compared to 1950*DA*. Therefore greater impact energies by *Eros* displace greater amount of ejecta which are thrown at higher velocities and therefore cover greater distances. Considering each object, increasing the initial density results in increasing impactor momentum at ground impact and hence greater deposition effects.

3.6. Airburst

Figures 35 and 36 show the peak overpressure values varying with distance from the impact site for objects 1950*DA* and *Eros* respectively. For similar initial object densities, pressure generated by *Eros* at a given distance is greater than that generated by 1950*DA* at a similar distance. *Eros* also generates pressures that reach far greater distances than those by 1950. Pressure by *Eros* also transitions from $1/r^{2.3}$ behavior to $1/r$ behavior at distances further than that of 1950*DA*. For a single object, increasing the initial object densities results in greater overpressure at a fixed distance. The difference between consecutive overpressures also increases with increasing initial object density. Transition from $1/r^{2.3}$ behavior to $1/r$ occurs at distances closer to the impact site for less dense impactors than for more dense impactors. In all cases, the rates of decrease of peak overpressure are similar except at the transition distances rx ranging from approximately 156598 to 314868 m for 1950DA and 1466594 m to 2934863 m for Eros. For both objects, the difference between consecutive pressure values decreases with increasing initial object density.

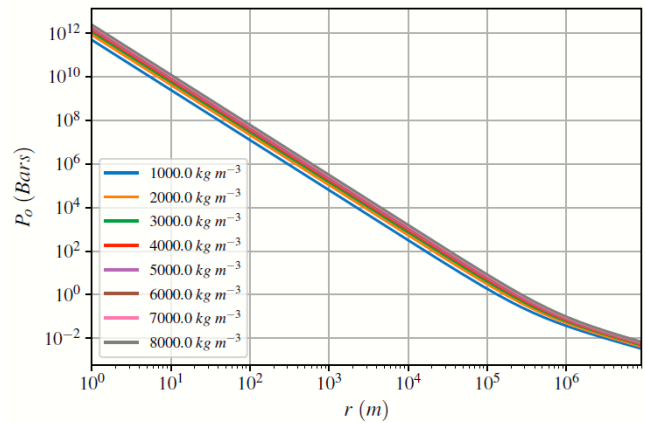

Figure 35: Peak Overpressure $P_o$ against distance $r$ plotted for Object (29075) 1950*DA* assigned different initial densities

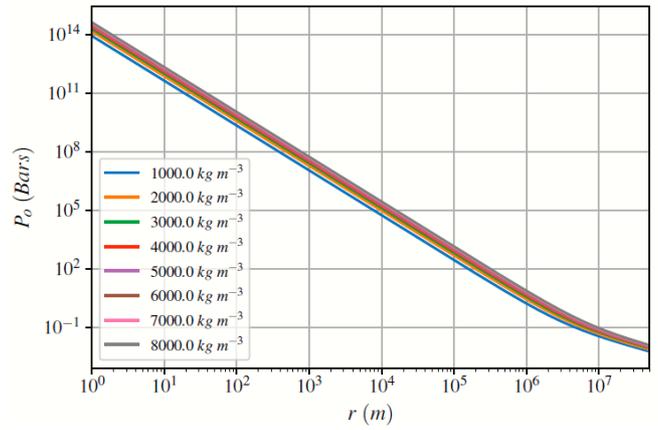

Figure 36: Peak Overpressure $P_o$ against distance $r$ plotted for Object 433 *Eros* assigned different initial densities

Figure 37 and 38 show the maximum wind velocities at different distances from the impact site for pressure generated by objects 1950*DA* and *Eros* respectively. For similar initial object densities, generated pressures by *Eros* have faster wind velocities than pressures generated by 1950*DA* at similar distances. Wind velocities generated by *Eros* also vary at far greater distances than wind velocities by 1950*DA*. Increasing the initial density of each object results in increase in wind velocities at fixed distances. However, the difference between consecutive wind velocities decreases with increasing initial object density. The rates of wind velocity decrease is similar for all object densities except at the transition distances.

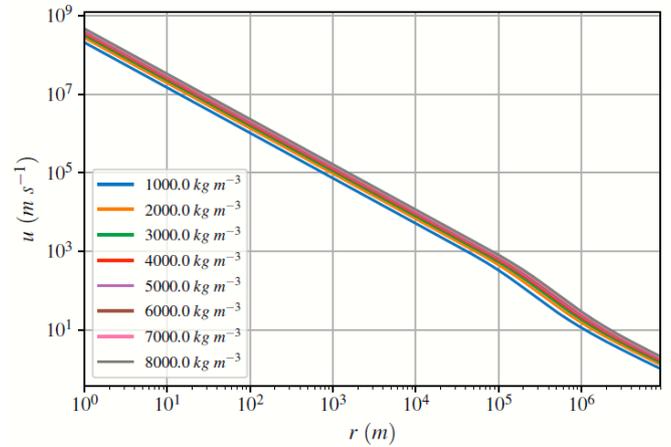

Figure 37: Maximum Wind Velocity $u$ against distance $r$ plotted for Object (29075) 1950*DA* assigned different initial densities

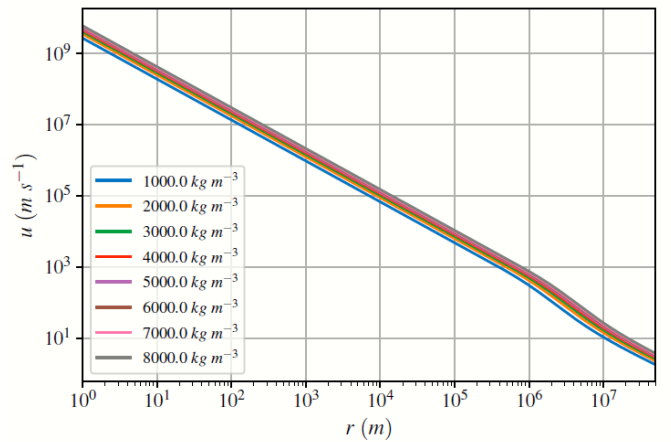

Figure 38: Maximum Wind Velocity $u$ against distance $r$ plotted for Object 433 *Eros* assigned different initial densities

Figures 39 and 40 show the arrival times at different distances from the impact site for blast waves generated by objects $1950DA$ and $Eros$ respectively. Given similar initial object densities, blast waves produced by $Eros$ have shorter arrival times than $1950DA$ at similar distances. Increasing the initial density of each object results in increase in shorter blast wave arrival times. The rates of increase in arrival times is similar for all object densities except at the transition distances.

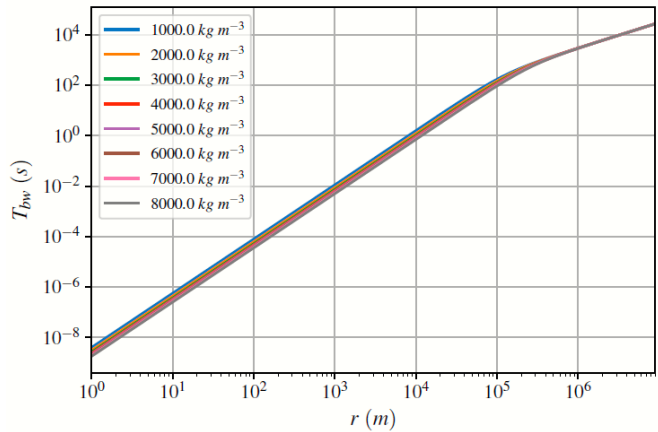

Figure 39: Blastwave arrival time $T_{bw}$ against distance $r$ plotted for Object (29075) $1950DA$ assigned different initial densities

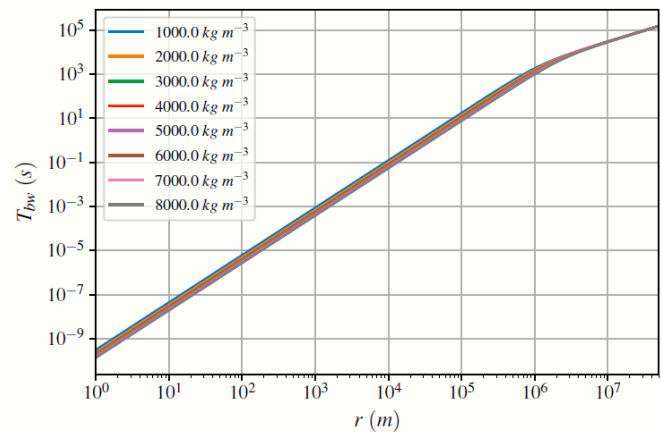

Figure 40: Blastwave arrival time $T_{bw}$ against distance $r$ plotted for Object 433 $Eros$ assigned different initial densities

Air burst effects depend to great extent on the energy of impact and the atmospheric conditions within which the events occur. Given similar atmospheric conditions, Eros still maintains greater kinetic energy after entry than $1950DA$. Therefore $Eros$ generates greater disturbance upon ground impact than $1950DA$, some of which is transferred to the atmosphere. Therefore the pressure above ambient atmospheric pressure is greater for $Eros$ than for $1950DA$. Greater pressures in the case of Eros result in higher wind velocities and faster arrival times than for $1950DA$. The same is true for a single object assigned different densities. Increasing the initial object density increases the momentum of impact by an object, thus resulting in generation of greater pressures with high wind velocities and shorter arrival times at similar distances.

## 4. Conclusion

The Large-Earth Impact Model used in this research was built on many assumptions and scaling techniques that reduce the resolution of the model. However it is convenient for pointing out key dynamics involved in Large Impact Events and providing a rough estimate of magnitudes of effects involved. The model describes the evolution several key parameters which include the impactor diameter, density and velocity. Impacts by massive Near-Earth asteroids like Eros are more resistant to disruption by the atmosphere and are highly likely to generate effects of the Chixculub scale. Impacts by low mass Asteroids like (29075) $1950DA$ are subjected to greater disruption upon atmospheric entry which reduces their initial kinetic energies resulting in effects of a comparably lower scale.

## 5. Acknowledgement

Funding: This research did not receive any specific grant from funding agencies in the public, commercial, or not-for-profit sectors.